\documentclass[12pt,preprint]{aastex}

\usepackage{amsmath}
\usepackage{amssymb}
\usepackage{xspace}
\usepackage{color}
\usepackage{ifthen}
\usepackage{xspace}

\slugcomment{Draft for ApJ, \today}

\usepackage{ulem}
\newcommand{\Msun}{{\ensuremath{\mathrm{M}_{\odot}}}\xspace}
\newcommand{\K}{{\ensuremath{\mathrm{K}}}\xspace}
\newcommand{\yr}{{\ensuremath{\mathrm{yr}}}\xspace}
\newcommand{\E}[1]{{\ensuremath{\times10^{#1}}}\xspace}
\newcommand{\erg}{{\ensuremath{\mathrm{erg}}}\xspace}
\newcommand{\B}[1]{{\ensuremath{\left(#1\right)}}\xspace}

\begin{document}

\title{Production of $^{26}$Al, $^{44}$Ti, and $^{60}$Fe in Core-Collapse Supernovae: Sensitivity to the Rates of the Triple Alpha
and $^{12}$C($\alpha,\gamma$)$^{16}$O Reactions}

\author{Clarisse~Tur\altaffilmark{1}}
\affil{National Superconducting Cyclotron Laboratory,\\
Michigan State University, 1 Cyclotron Laboratory, East Lansing, MI 48824-1321\\
Joint Institute for Nuclear Astrophysics}
\email{tur@nscl.msu.edu}

\author{Alexander~Heger\altaffilmark{1}}
\affil{School of Physics and Astronomy,
University of Minnesota, Twin Cities,
Minneapolis, MN 55455-0149; \\
Nuclear and Particle Physics, Astrophysics and Cosmology Group,
T-2, MS B227,
Los Alamos National Laboratory,
Los Alamos, NM 87545}
\email{alex@physics.umn.edu}

\author{Sam~M.~Austin\altaffilmark{1}}
\affil{National Superconducting Cyclotron Laboratory,\\
Michigan State University, 1 Cyclotron Laboratory, East Lansing, MI 48824-1321\\
Joint Institute for Nuclear Astrophysics\phantom{.}}
\email{austin@nscl.msu.edu}

\altaffiltext{1}{Joint Institute for Nuclear Astrophysics}

\begin{abstract}
\noindent
We have studied the sensitivity to variations in the the triple alpha
and $^{12}$C($\alpha,\gamma$)$^{16}$O reaction rates of the production
of $^{26}$Al, $^{44}$Ti, and $^{60}$Fe in core-collapse supernovae.
We used the KEPLER code to model the evolution of $15\,\Msun$,
$20\,\Msun$, and $25\,\Msun$ stars to the onset of core collapse and
simulated the ensuing supernova explosion using a piston model for the
explosion and an explosion energy of $1.2\,\E{51}\,\erg$.
Calculations were performed for the Anders and Grevesse (1989) and
Lodders (2003) abundances.  Over a range of twice the experimental
uncertainty, $\sigma$, for each helium-burning rate the production of
$^{26}$Al, $^{60}$Fe, and their ratio vary by factors of five or more.
For some species, similar variations were observed for much smaller
rate changes, $0.5\,\sigma$ or less.  The production of $^{44}$Ti was
less sensitive to changes in the helium-burning rates.  Production of
all three isotopes depended on the solar abundance set used for the
initial stellar composition.
\end{abstract}


\section{Introduction}

Observations of gamma rays from radioactive nuclei, and, in
particular, from $^{26}$Al, $^{44}$Ti, and $^{60}$Fe, with halflives of
$7.2\times 10^5\,\yr$, $60\,\yr$, and $2.62\times 10^6\,\yr$
(\citealt{rug09}), respectively, provide information about the sites
and nature of stellar nucleosynthesis that is difficult to obtain in
other ways.  For example, the spatial distribution of $^{26}$Al
inferred from observation of its gamma rays indicates that their most
probable source is massive stars; a comparison with the observed
intensity of $^{60}$Fe, $^{26}$Al, and $^{44}$Ti gamma rays provides a
stringent test of supernova models; a comparison of the observed
intensity of the $^{26}$Al gamma rays and the calculated production of
$^{26}$Al in core-collapse supernovae provides an estimate of the rate
of supernovae in the galaxy; the failure to observe $^{44}$Ti gamma
rays, other than from Cas~A may indicate that emission of $^{44}$Ti in
supernovae is fairly rare, or that the supernova rate has been
overestimated.  For a discussion of these issues see \cite{lei09,die06a}.

Observations have now yielded accurate gamma ray intensities for these
nuclei (\citealt{die06a} and the references therein; \citealt{wan07}),
but the reliability of deductions based on these intensities is
limited by our understanding of the production mechanisms that
synthesize $^{26}$Al, $^{44}$Ti, and $^{60}$Fe.  For example,
estimates of the ratio $^{60}$Fe/$^{26}$Al have varied greatly as
discussed by \citet{lim06,woo07}.  Early calculations by
\citet{tim95} are in reasonable agreement with present results,
whereas some later calculations using the same supernova code yield
much different results.  A detailed discussion, tracing these changes
to the effects of reaction rate choices, can be found in
\citet{woo07}.  Additional uncertainties arising from different
astrophysical model assumptions or approximations are discussed in
\citet{woo07,lim06,you07}.  In addition, it appears that yields may be
significantly different when simulations are done in two or three
dimensions \citep{mea06}.

The yield uncertainties resulting from reaction rate uncertainties
that have been examined previously are mostly for reactions directly
involved in the nucleosynthesis processes, but other reaction rate
uncertainties may also be important.  In their recent study of
$^{26}$Al and $^{60}$Fe yields from massive stars \cite{lim06} showed
that the amount of C fuel left by the standard helium burning affects
all subsequent shell burning episodes and the yields of $^{26}$Al and
$^{60}$Fe. Since the yields of $^{12}$C and $^{16}$O at the completion
of core helium burning are determined by the rates for the
triple-alpha ($R_{3\alpha}$) and $^{12}$C($\alpha,\gamma$)$^{16}$O
($R_{\alpha,12}$) reactions, we expect the yields of $^{26}$Al and
$^{60}$Fe to depend on these rates. We have previously shown that the
production factors for many isotopes with $A<90$ depend strongly on
both $R_{3\alpha}$ and $R_{\alpha,12}$ \citep{tur07,tur09}.

There have been several studies of the effects of uncertainties in
these helium burning rates.  Only a few have involved changes in
$R_{3\alpha}$, mostly for studies of the effects of changes in
coupling constants over long time scales.  For example, \citet{sch04}
investigated the effects of changes in the excitation energy of the
Hoyle state at $7.65\,$MeV in $^{12}$C but considered only large
changes in rate and did not give yields for $^{60}$Fe, $^{26}$Al, and
$^{44}$Ti.  Other authors have discussed the results of changes in
$R_{\alpha,12}$ but yields of $^{60}$Fe, $^{26}$Al, and $^{44}$Ti are
not often given. \cite{imb01} evolved a $25\,\Msun$ star at two values
of $R_{\alpha,12}$ which differ by a ratio of $2.35$, corresponding to
about $0.7$ and $1.6$ on the rate scales we use for $R_{\alpha,12}$
if one normalizes to the value of the rate at $300\,$keV. We will show a comparison with the \citeauthor{imb01} results in Section~3. \cite{hof99} used two different $R_{\alpha,12}$ and present results for a $15\,\Msun$ star, but also
changed many other reaction rates so an accurate comparison with our
results is not possible.

It is clear that available studies do not provide a detailed
estimate of the variation of the yields of $^{60}$Fe, $^{26}$Al, and
$^{44}$Ti with changes in the helium burning rates.  This shortcoming
is especially important because one anticipates that the variation of
the yields with rates might be rapid.  The lifetimes of the long lived
isotopes depend strongly on the temperature of their environment.  For
example, \citet{lim06}, note that the lifetime of $^{26}$Al drops to
$0.19\,\yr$ at $\log\B{T/\K}=8.4$ and that of $^{60}$Fe to $0.5\,\yr$
at $\log \B{T/\K}=9$.  Their yields will then depend on whether or not
they are formed in a convective region of the star and are carried
relatively quickly to cooler regions.  Since the details of the
convective structure of a star depend on the helium burning rates, we
expect that the production of $^{60}$Fe, $^{26}$Al, and $^{44}$Ti
might show a similar sensitivity.  A detailed study, with closely
spaced rate changes is necessary.

In this paper we describe a systematic study of the changes in
supernovae synthesis of $^{26}$Al, $^{44}$Ti, and $^{60}$Fe that
result from changes in the helium burning reaction rates and changes
in the initial stellar abundances.  The emphasis is not on the
\emph{absolute} yields, but on the \emph{variations} in the yields
induced by changes in the helium burning reaction rates and the
initial abundances.  In total about 200 stars were evolved; the
specific values of the helium burning rates used are shown in Fig.~1
of \cite{tur07}.  We find that over the $\pm 2\sigma$ range of rates,
the yield changes are large, often a factor of five or more, and that
the yields of $^{60}$Fe, and $^{26}$Al often vary rapidly with
reaction rate and do not show simple monotonic behavior.

In \S2 we present details of the calculations.  In \S3, we compare our
results for the standard values of the helium burning reaction rates
to prior calculations (AG89 abundances), and then present the outcome
of our rate and abundance changes for the three radioactivities under
study.  In \S4, we discuss other important uncertainties that can
impact the yield predictions and compare our results with results of
observations.  Our conclusions are given in \S5.

\section{Stellar models and reaction rates}

We used the KEPLER code \citep{wea78,woo95,rau02,woo02} to simulate
stellar evolution from standard hydrogen burning up to core-collapse;
a piston placed at the base of the oxygen shell was then used to
simulate the explosion.  The outward velocity of the piston was set to
impart a kinetic energy of $1.2\E{51}\,\erg$ to the ejecta as
they escape to infinity.  After estimating the fallback from the
hydrodynamic supernova simulations, the final yields were determined
by employing the parametrization of the mixing that was used by
\citet{woo07}.  See \citet{tur07} for a more detailed description of
the models.

We used the rates of \cite{rau00} for the $^{26}$Mg($p,n$),
$^{26}$Al($n,p$), $^{26}$Al($n,\alpha$), $^{40}$Ca($\alpha,\gamma$),
$^{44}$Ti($\alpha,\gamma$), $^{44}$Ti($\alpha,p$),
$^{59}$Fe($n,\gamma$), and $^{60}$Fe($n,\gamma$) reactions, the rates
of \cite{ili01} for the $^{25}$Mg($p,\gamma$), and
$^{26}$Mg($p,\gamma$) reactions, the rates of \cite{fis01} for the
$^{44}$Ti($\alpha,p$) reaction, and for $^{22}$Ne($\alpha,n$) rate we
used the lower limit from \citep{jae01}.  The calculations described
in this paper were performed over a significant time period.
Sometimes the preferred rates changed over that period, but because of
our desire for a systematic set of studies, focusing on the rates of
the helium burning reactions, we retained our initial choices for the
entire set of simulations. We expect that our simulations will provide
a reasonable estimate of the \emph{relative} yield uncertainties
resulting from the uncertainties in the helium burning rates, even if
other rates or approximations change.  A good overview over the
reaction rates used can also be found in \citet{rau02}.

We calculated $^{26}$Al, $^{44}$Ti and $^{60}$Fe yields for $\pm
2\,\sigma$ variations of $R_{3\alpha}$ and $R_{\alpha,12}$ from their
standard values as given by \citet{cau88} and by $1.2$ times the rate
recommended by \citet{buc96}, respectively.  The $\pm 1\,\sigma$
experimental uncertainties were taken to be $12\,\%$ and $25\,\%$.  We
performed calculations for $15\,\Msun$, $20\,\Msun$, and $25\,\Msun$
stars and for three different rate sets: (1) $R_{3\alpha}$ was kept
constant at its value from \citet{cau88}, and $R_{\alpha,12}$ was
varied; (2) $R_{\alpha,12}$ was held constant at $1.2$ times the rate
recommended by \citet{buc96} and $R_{3\alpha}$ was varied; (3) both
standard values of the rates were varied by the same factor, so their
ratio remained constant, to investigate the extent to which only the
ratio of the two rates is important.

We also determined the yields for two different stellar compositions:
\citet{lod03} and \citet{and89}, hereafter L03 and AG89. The L03
abundances for C, N, O, Ne are rather close to those of \citet{asp09}.

\section{Results}

\subsection{Comparison with previous calculations}

In Table~\ref{tbl-1}, we show the final yields for $^{26}$Al,
$^{44}$Ti, and $^{60}$Fe for our $15\,\Msun$, $20\,\Msun$, and
$25\,\Msun$ stars, and their average using a Scalo \citep{sca86}
Initial Mass Function (IMF) with a slope of $\gamma=-2.6$, for the
standard values of $R_{3\alpha}$ and $R_{\alpha,12}$, and for both the
AG89 and the L03 abundances.  The difference in the yields due to the
initial composition is largest for the $25\,\Msun$ star, where the
yields are reduced by $\sim28\,\%$, $\sim15\,\%$, and $\sim81\,\%$ for
$^{26}$Al, $^{44}$Ti, and $^{60}$Fe, respectively, in going from the
AG89 to the L03 abundances.  This effect is significant, but smaller
than the effect of variations in the helium burning rates, as we shall
see in the next subsection.  The major difference in the abundance
sets is the substantially reduced CNO content of L03 compared to AG89
(see Fig.~2 in \citealt{tur07}).  The central carbon mass fraction for
the $25\,\Msun$ star, however, changes only by about $5\,\%$ and is
$0.1903$ ($0.1790$) for the AG89 (L03) abundances \citep{tur07}.  The
IMF-averaged values for $^{26}$Al and $^{44}$Ti are very similar for
the two abundance sets.

\begin{figure*}
\centering
\includegraphics[angle=90,width=0.475\textwidth]{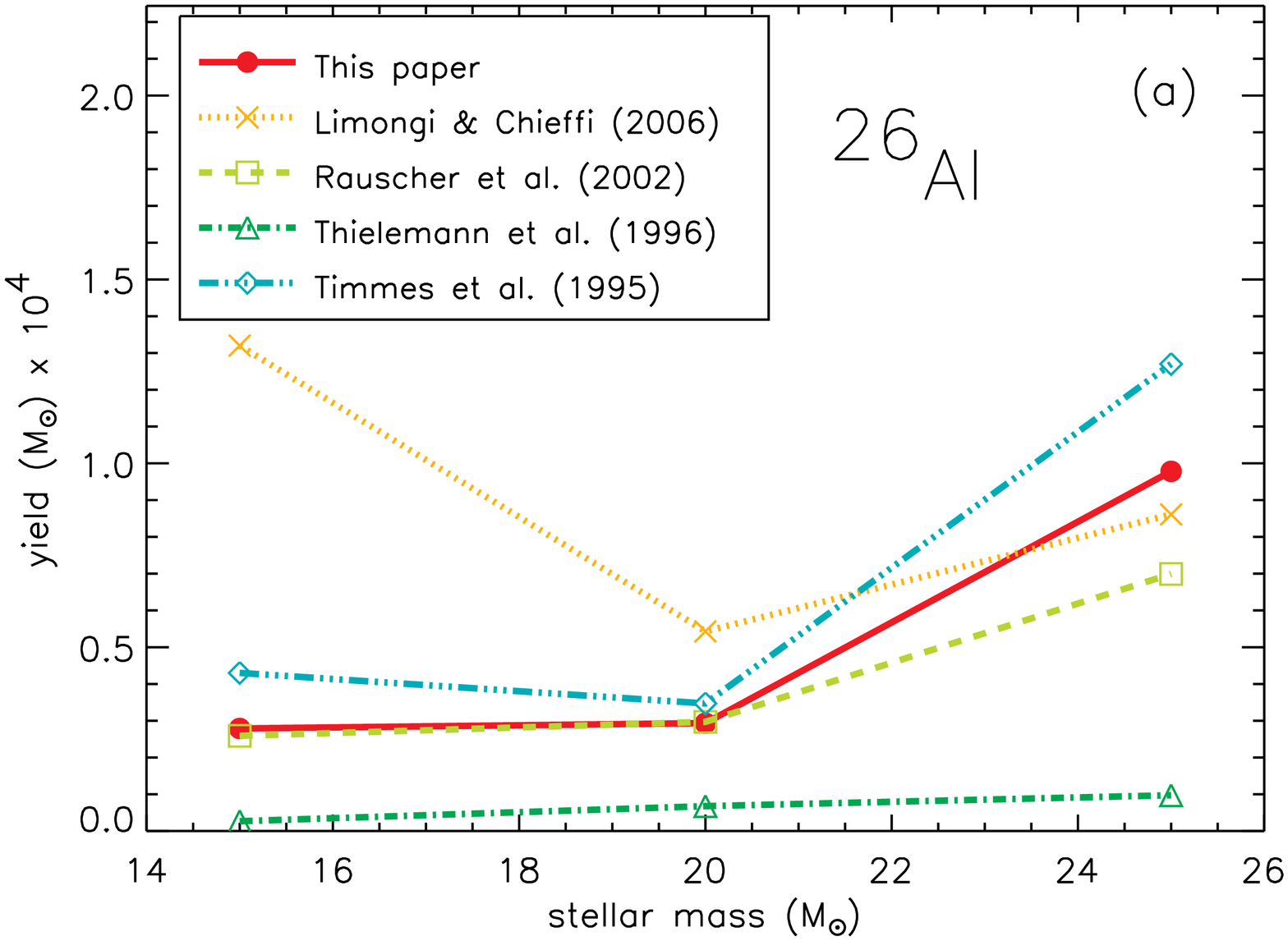}
\hfill
\includegraphics[angle=90,width=0.475\textwidth]{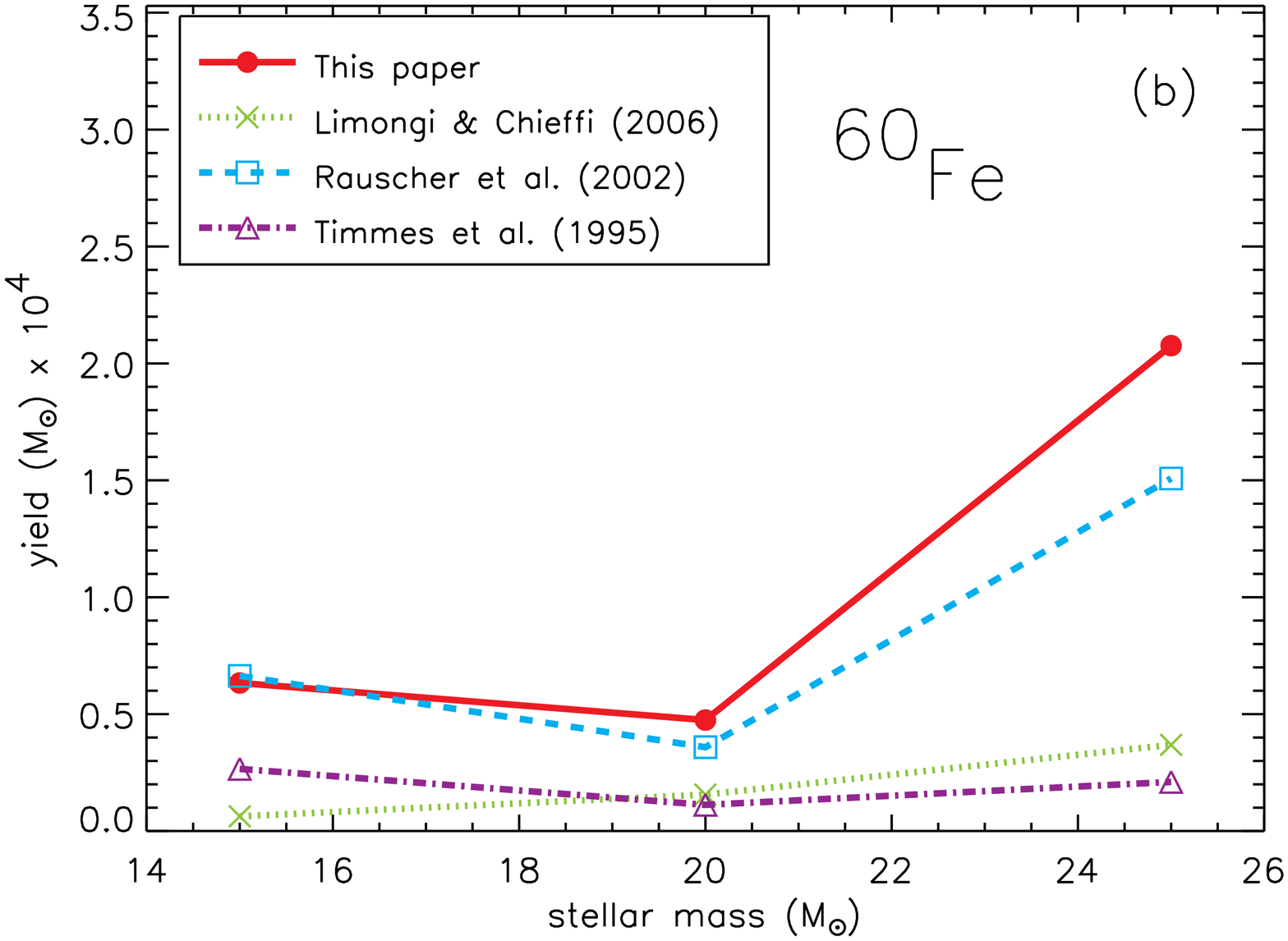}
\hfill
\includegraphics[angle=90,width=0.475\textwidth]{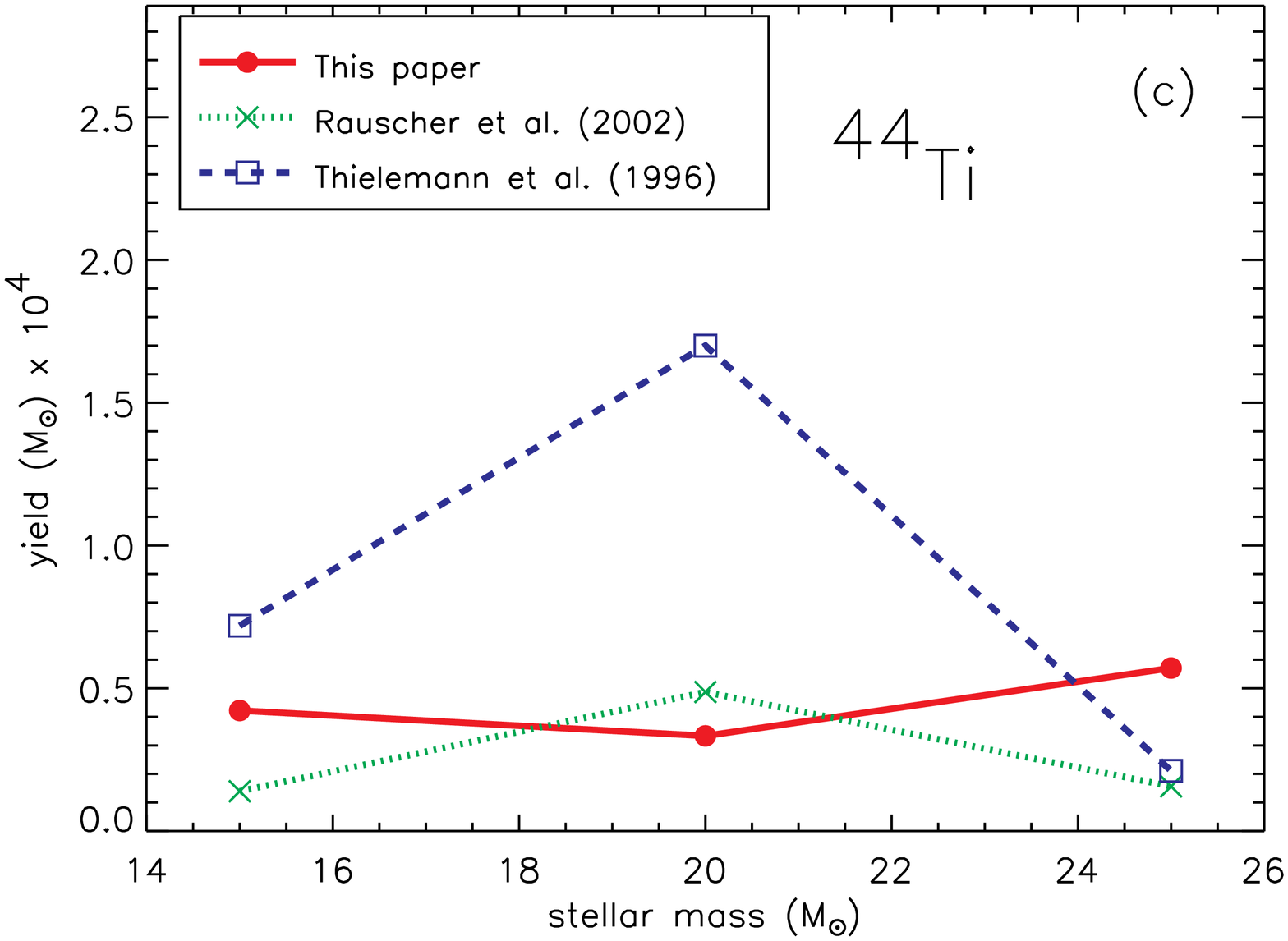}
\caption{Yields versus stellar mass: comparison between the present
  study (standard values of the helium burning reaction rates, AG89
  abundances) and previous publications. \textbf{(a)} $^{26}$Al.
  \textbf{(b)} $^{60}$Fe. \textbf{(c)} $^{44}$Ti.}
\label{fig1}
\end{figure*}

Figure~\ref{fig1} shows a comparison of our results for the yields of
$^{26}$Al, $^{44}$Ti, and $^{60}$Fe (for standard values of
$R_{3\alpha}$ and $R_{\alpha,12}$, and the AG89 abundances), and the
results from some recent calculations: in Figure~\ref{fig1}\emph{a}
our $^{26}$Al yields are shown together with those of
\citet{lim06,rau02,thi96,tim95} (the latter is based on the survey by
\citealt{woo95}); in Figure~\ref{fig1}\emph{b}, we compare our
$^{60}$Fe yields to those of \citet{lim06,rau02,tim95}; in
Figure~\ref{fig1}\emph{c}, the $^{44}$Ti yields are plotted for our
study as well as for \citet{rau02,thi96}.  The \citet{thi96} yields
for $^{60}$Fe are not shown because they were much smaller than the
yields shown here.  \citet{lim06,tim95} only studied the yields for
$^{26}$Al and $^{60}$Fe, so no $^{44}$Ti yield can be shown.

The $^{26}$Al yields found for most of these studies are in rough
agreement, given the uncertainties due to reaction rate choices and
astrophysical model differences described in the Introduction.  The
larger differences for $^{60}$Fe can mainly be explained by the
reaction rate differences discussed in \cite{woo07}.  The yield curves
of Figure~\ref{fig1} for the present study have dips at $20\,\Msun$
for the three isotopes studied in this paper.  This dip corresponds to
an inversion of the monotonically increasing general trend in the
mass-radius relation as noted by \cite{lim06}.  A related behavior for
the $20\,\Msun$ star was observed for medium-weight isotopes
\citep{tur07,rau02}, as a result of merging of the O, Ne,
and C burning shells about a day prior to the supernova explosion.

\subsection{Effect of variations of the helium burning reaction rates}

We adopt a three-character notation to label our plots, e.g., LA3,
similar to the notation adopted by \cite{tur07}.  The first character
can be an ``L'' (to denote the L03 initial abundances) or an ``A''
(for the AG89 initial abundances).  The second character denotes the
study: ``A'', when $R_{3\alpha}$ was kept constant, and
$R_{\alpha,12}$ was varied, ``B'', when both rates were varied by the
same factor, so their ratio remained constant, and ``C'', when
$R_{\alpha,12}$ was held constant, and $R_{3\alpha}$ was varied.  The
third character is the number of stars included in the IMF average; it
is typically 3, for averages over the three stars ($15\,\Msun$,
$20\,\Msun$, and $25\,\Msun$).  When no third character is present the
values apply to a single star.

\begin{figure*}
\centering
\includegraphics[angle=90,width=0.475\textwidth]{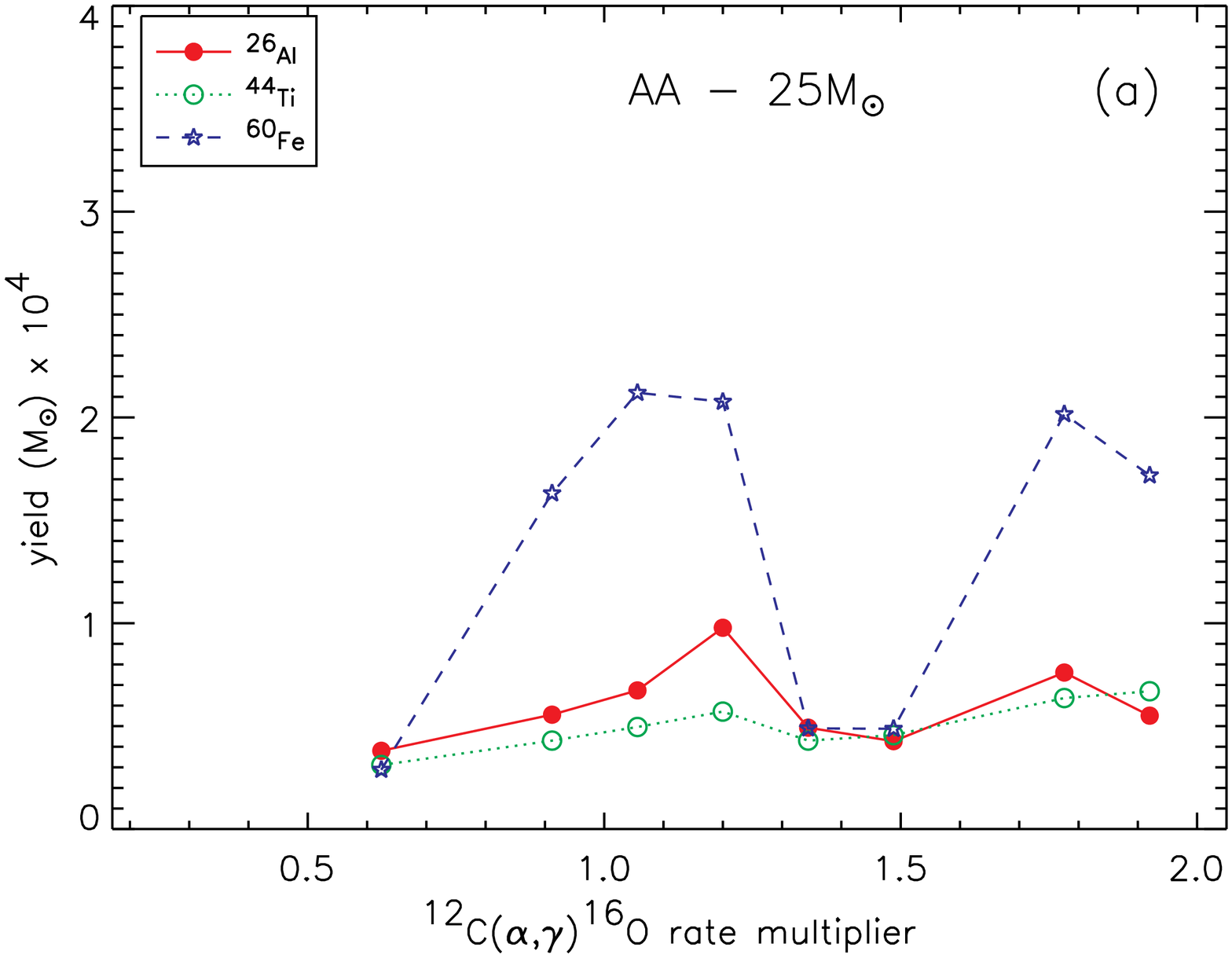}
\hfill
\includegraphics[angle=90,width=0.475\textwidth]{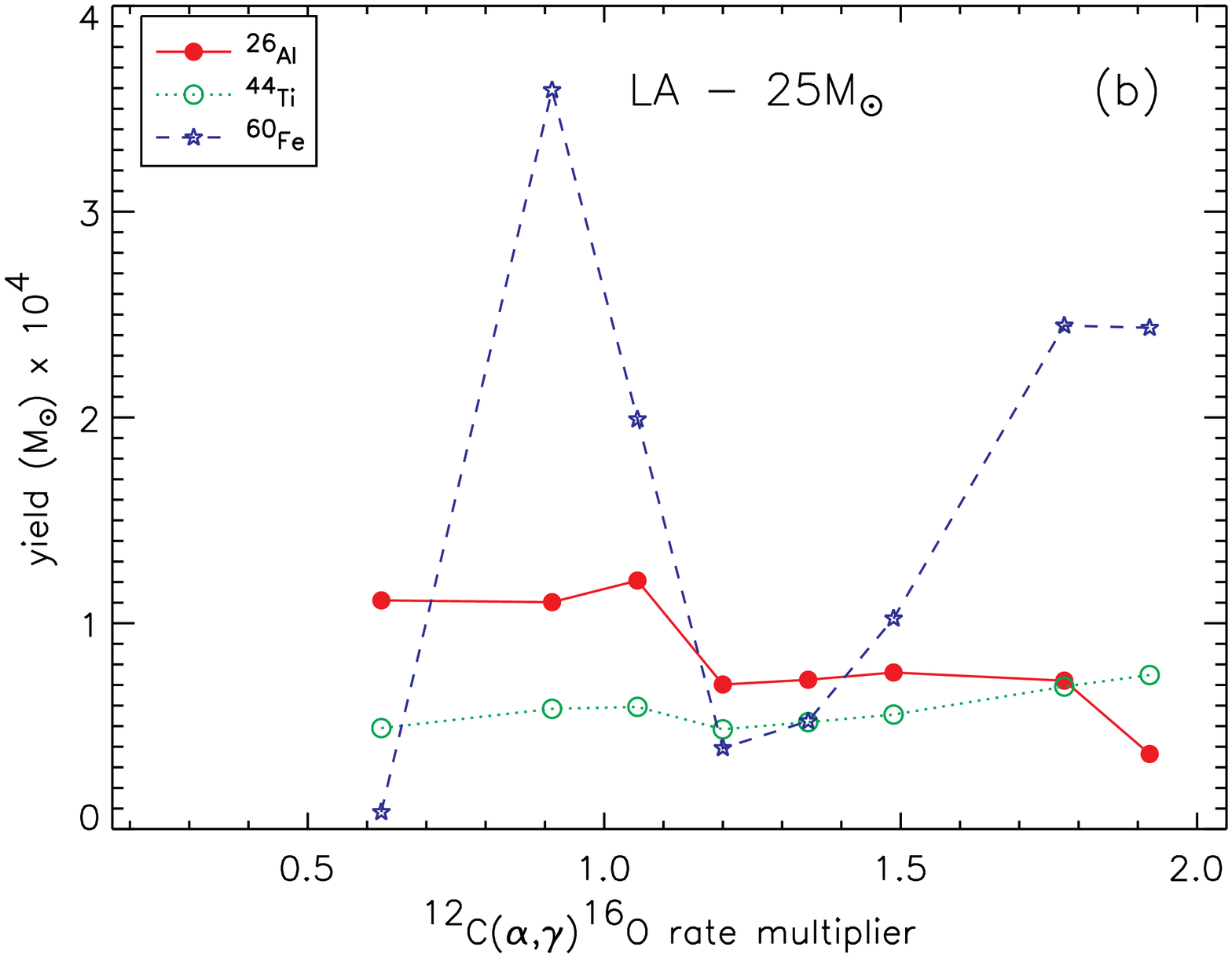}
\hfill
\includegraphics[angle=90,width=0.475\textwidth]{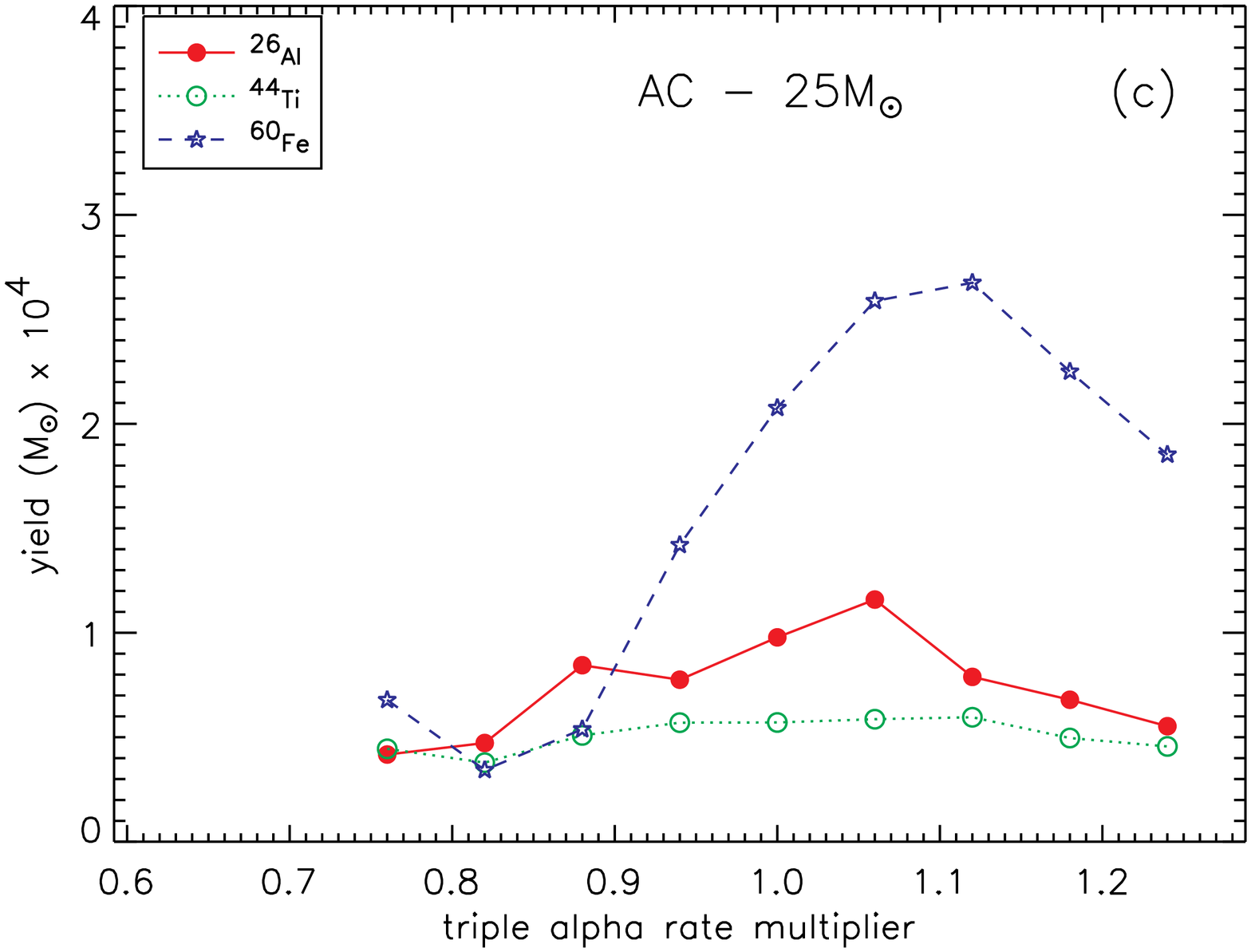}
\hfill
\includegraphics[angle=90,width=0.475\textwidth]{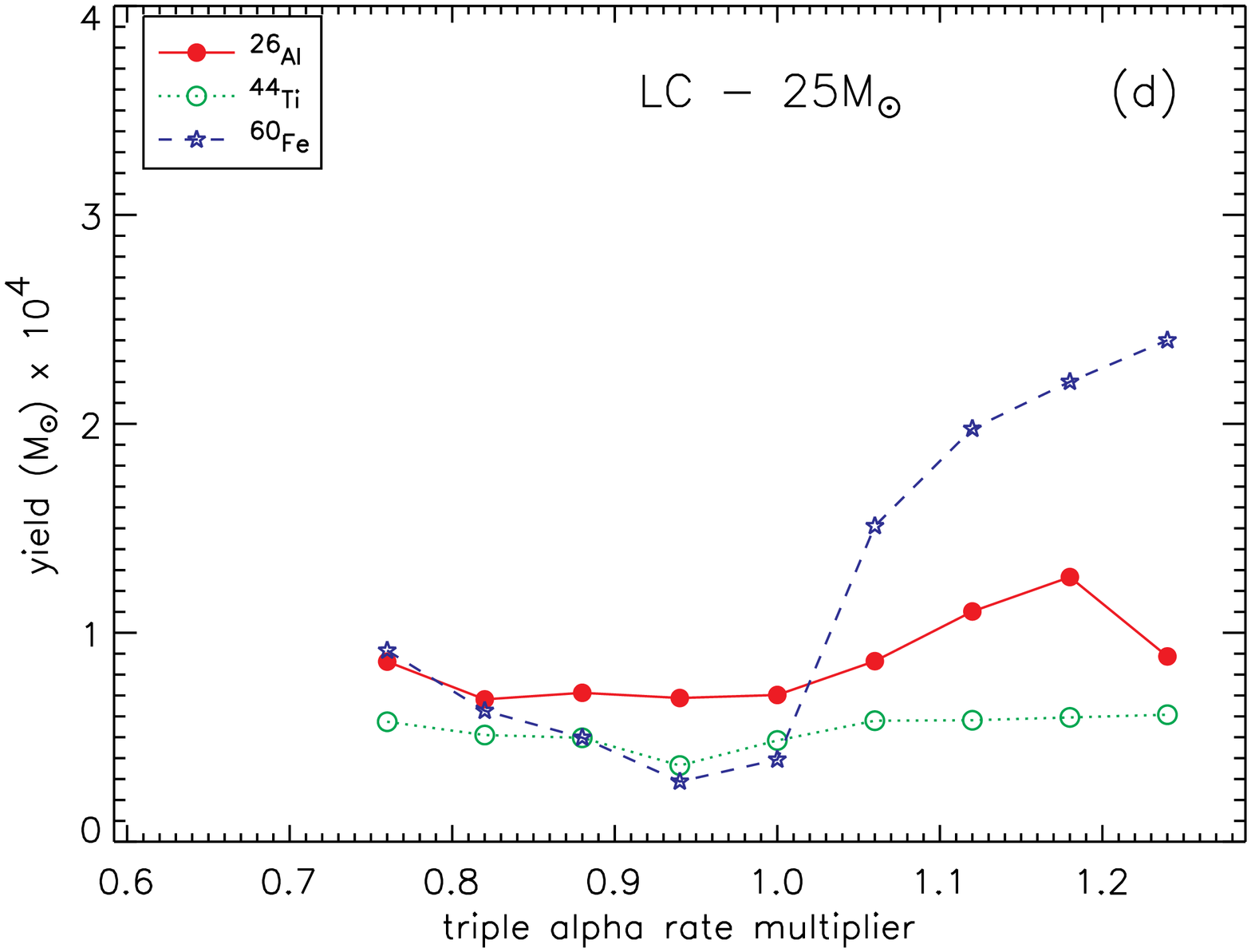}
\hfill
\includegraphics[angle=90,width=0.475\textwidth]{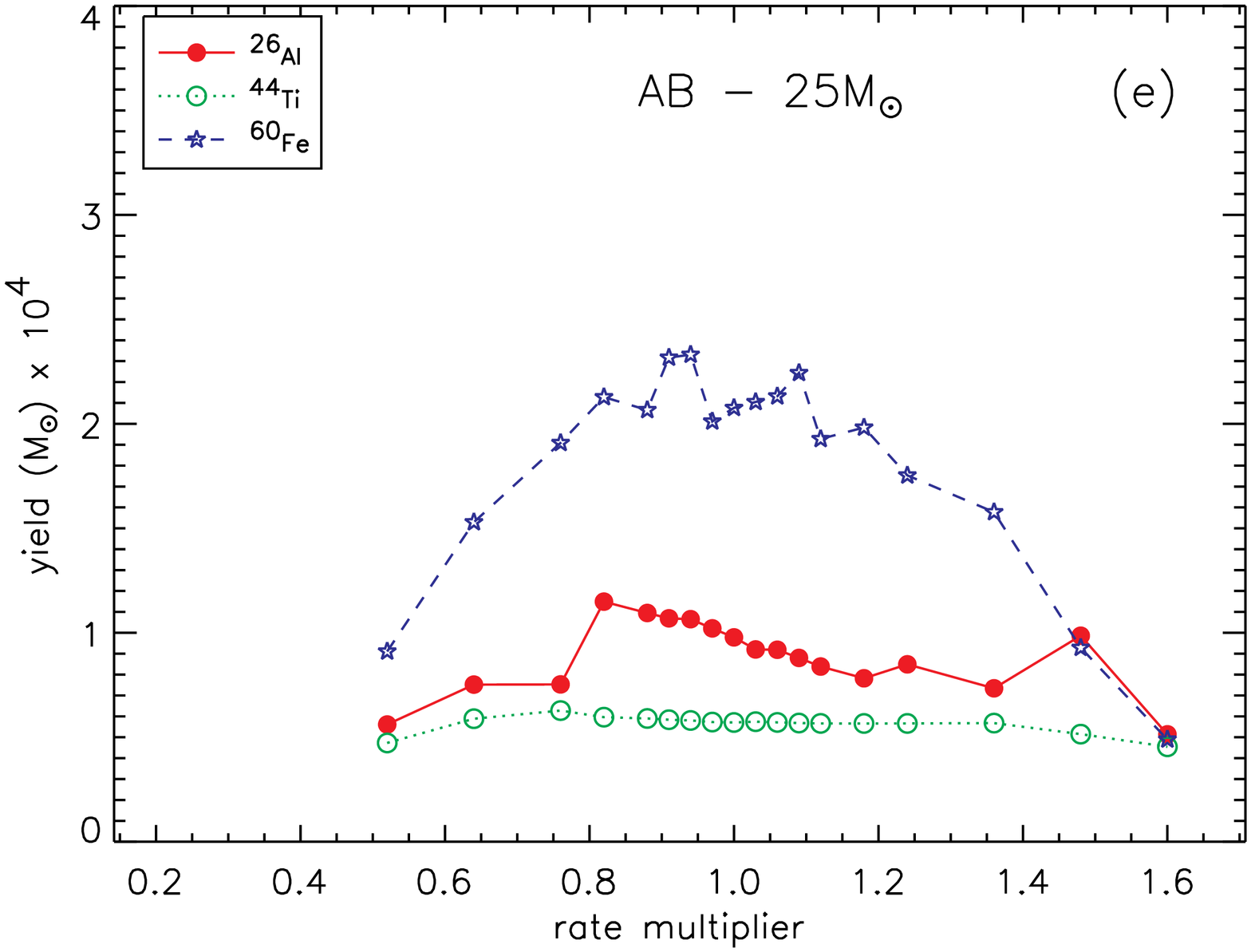}
\hfill
\includegraphics[angle=90,width=0.475\textwidth]{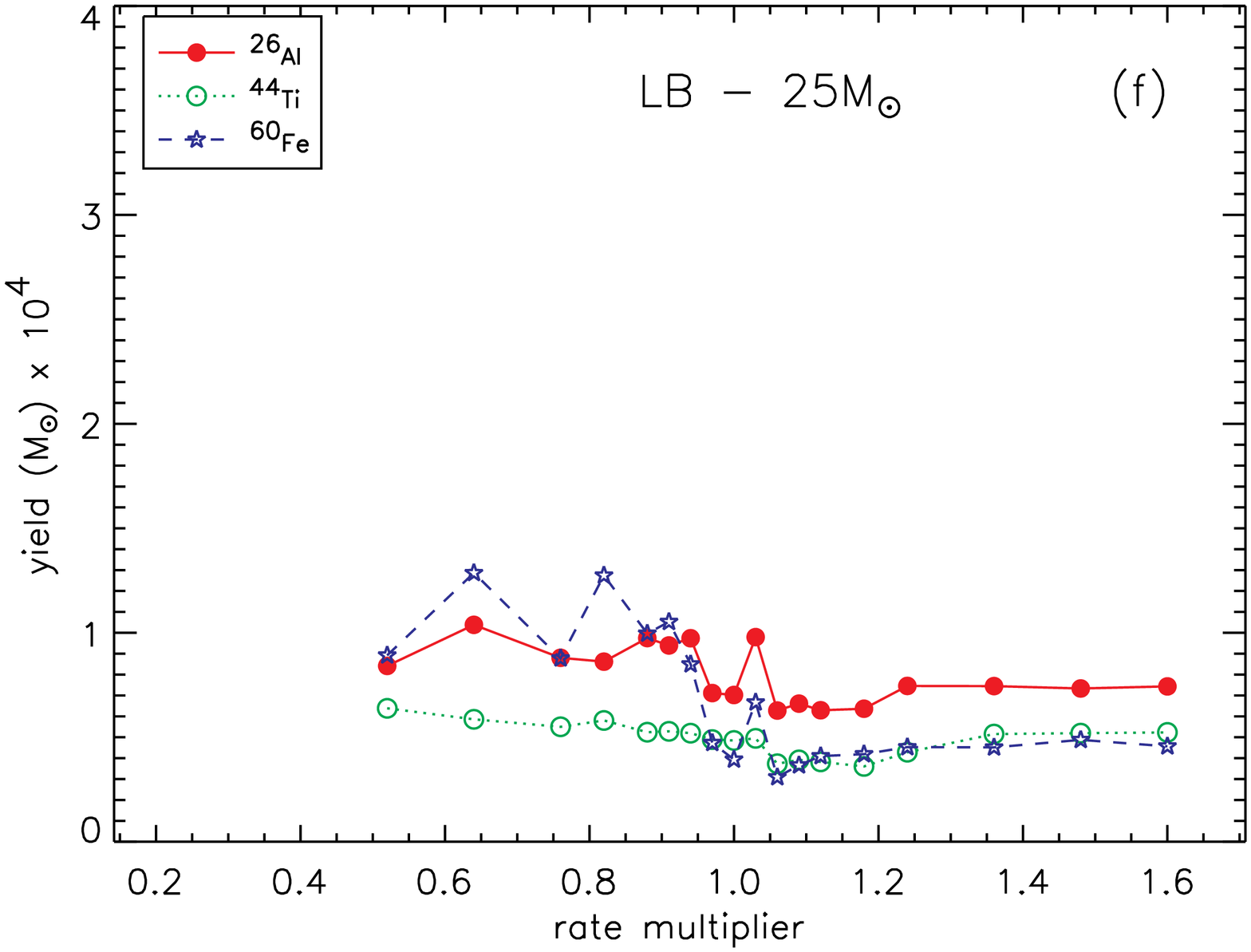}
\caption{Yields versus reaction rate for the $25\,\Msun$ model.  For
  details see the text.  \textbf{(a)} AA series. \textbf{(b)} LA
  series. \textbf{(c)} AC series.  \textbf{(d)} LC
  series. \textbf{(e)} AB series. \textbf{(f)} LB series.}
\label{fig2}
\end{figure*}

Figures~\ref{fig2}\emph{a}-\emph{f} show the yields of $^{26}$Al,
$^{44}$Ti, and $^{60}$Fe for the $25\,\Msun$ star as a function of
reaction rate for the three sets of rates (see Introduction) and for
the two abundance sets we used: AG89 and L03.
Figures~\ref{fig3}\emph{a}-\emph{f} show the same for the IMF-averaged
yields.  The yield often varies non-monotonically and sharply with
rate.  We find a strong dependence of the yields on $R_{3\alpha}$ and
$R_{\alpha,12}$, and a smaller dependence on the initial solar
abundances.  The strongest variations occur for the $^{60}$Fe yields,
with weaker variation for the $^{26}$Al yields, and still weaker
variations for $^{44}$Ti.

For the L03 abundances, in the LA series for the $25\,\Msun$ star, the
$^{60}$Fe yield reaches a maximum value of $3.590\E{-4}\,\Msun$ when
$R_{\alpha,12}$ is reduced by $24\,\%$ (relative to its standard
value), and a minimum of $8.285\E{-6}\,\Msun$ when it is reduced by
$48\,\%$, a factor of $\sim43$ change in yield.  Changes in
$R_{3\alpha}$ can cause similarly large effects: in the LC series for
the $25\,\Msun$ star, the $^{60}$Fe yield is at it maximum value of
$2.400\E{-4}\,\Msun$ when $R_{3\alpha}$ is increased by $24\,\%$
(relative to its standard value), and a minimum value of
$2.881\E{-5}\,\Msun$ when $R_{3\alpha}$ is reduced by $6\,\%$, a factor
of $\sim12$ change. Even for $^{44}$Ti, differences of a factor of
about $2$ are observed for certain cases.  The differences for the IMF
averaged case are somewhat smaller but still substantial, especially
for $^{60}$Fe.

For the AB and LB series, where both rates are a given multiple of
their standard values, the variations are weaker. The variation of the
central carbon abundances with reaction rate is about one-third of
that for comparable variation of the reaction rate $R_{\alpha,12}$
alone (\cite{tur07}). This is consistent with the generally weaker
variations in the yields of the radioactive nuclides.

Since the half-lives of $^{26}$Al and $^{60}$Fe ($\sim10^{6}\,\yr$) are
much longer than the average time between two galactic SNII events
($\sim2$ per century), a steady state assumption, where the production
rate balances the decay rate is reasonable.  The changes in the
production yields discussed above then translate directly into changes
in the predicted photon flux.

More detailed results for variation of $R_{3\alpha}$ and
$R_{\alpha,12}$, and for the two initial abundance sets are given in
Tables~\ref{tbl-2} and \ref{tbl-3}.

We can compare our results to those of \cite{imb01} at
$R_{\alpha,12}=0.7$ and $1.6$.  Since we did not do simulations at
these exact values, we have read values from
Fig.~\ref{fig2}\emph{a}. The results, especially for $^{60}$Fe, are
subject to large extrapolation uncertainties.  The energy dependencies of the reaction rates also differ. The ratio of our
results at $0.7$ to those at $1.6$ for $^{26}$Al ($^{60}$Fe) is $0.8$
($0.6$) and those of \citeauthor{imb01} is $0.17$
($1.22$). Individual values of the yields agree within a factor of
$2-3$.  It is difficult to judge whether these differences are
significant, given the different assumptions on stellar physics that
entered the calculations, as outlined in the introduction.

\subsection{Production mechanisms}

Figure~\ref{fig4} shows the yields of $^{26}$Al, $^{44}$Ti, and
$^{60}$Fe for a $25\,\Msun$ star using the L03 abundances at
various stages of stellar evolution for the standard values of
$R_{3\alpha}$ and $R_{\alpha,12}$ (Figure~\ref{fig4}\emph{a}) and when
$R_{3\alpha}$ is increased by $18\,\%$, $R_{\alpha,12}$ maintained at
its standard value (Figure~\ref{fig4}\emph{b}). The stellar burning
stages for which yields are plotted and that are also used in
Figure~\ref{fig5} are:

\begin{itemize}
\item [\emph{He-ign:}] At this time $1\,\%$ of the helium has burnt in the
  center.

\item [\emph{He-dep:}] Just before central He depletion
  (when the central He mass fraction has reached $1\,\%$).

\item [\emph{C-ign:}] Just before central C ignition
  (when the central temperature reaches a value of $5\E8\,\K$).
  Central helium burning has finished.

\item [\emph{C-dep:}]  Central C depletion (when the
 central temperature has reached $1.2\E9\,\K$).  Central C
 burning contributes mostly to those regions of the star below the
 mass cut.

\item [\emph{O-dep:}] At central O depletion (when the
  central O mass fraction has dropped below $5\,\%$).

\item [\emph{Si-dep:}] Central Si depletion (when the central Si mass fraction
  drops below $10^{-4}$).

\item [\emph{pre-SN:}] At the pre-supernova stage when the contraction
  speed in the iron core reaches $1000\,$km$\,$s$^{-1}$.  This is
  typically about $0.25\,$s before core bounce.

\item [\emph{SN unmix:}] $100\,$s after the passage of the shock wave.
  At this time all thermonuclear reactions and neutrino-process have
  essentially ceased, but Rayleigh-Taylor (RT) instabilities that
  cause mixing after the passage of the shock have not yet mixed the
  ejecta (in our approximation).  This is only shown in
  Figures~\ref{fig5}\emph{d}, \ref{fig5}\emph{e}.

\item [\emph{Final:}] $>100\,$s after the passage of the shock wave.
  This includes mixing due to RT instabilities parametrized as in
  \citep{rau02} and fallback (see also \citealt{you07}).  In
  Figures~\ref{fig5}\emph{e,f} this has been labeled as ``\emph{SN
    mix}''.

\end{itemize}

The convective history for the above two cases is shown in
Figure~\ref{fig5}\emph{a,b} and the times of the snap-shots are
indicated.  Figure~\ref{fig5}\emph{c,d} shows the pre-supernova
structure for these two cases including convective regions and mass
fractions of $^{26}$Al, $^{44}$Ti, and $^{60}$Fe.  The inner dark gray
region indicates iron core where we no longer follow detailed
nucleosynthesis as these regions become part of the collapsing iron
core and disappear in the remnant.  Figure~\ref{fig5}\emph{e,f} shows
the distribution of the isotopes inside the star at the different
evolution stages.  Here, light gray regions indicate where detailed
nucleosynthesis was no longer followed.  Notably, in both of the last
two rows of Figure~\ref{fig5} the vertical solid black line indicates
the location of the ``mass cut'' after the supernova explosion -
everything below that line disappears in the remnant.  The solid lines
and symbols in Figure~\ref{fig4} show only the contributions outside
the this mass cut, the dashed line and hollow symbols show mass
contained in the entire star as computed at each stage.  In general,
the piston in the SN explosion can be below the mass cut if there is
``fallback'' -- mass that has first moved outward during the SN
explosion, but then did not escape and fell back onto the compact
remnant instead; in the specific cases shown here there was not such
``fallback''.

As is shown in Figure~\ref{fig5}\emph{a,b} changes in reaction rates
can cause significant differences in the convective history of a star.
Convective zones can carry radioactive nuclei to cooler zones of the
star where their effective lifetime is longer and their survival is
enhanced.  It can bring nuclei to hot regions where they can be
destroyed, or transport ``fuel'' to very hot regions to synthesize
nuclei.  In our case we find a \textrm{convective} oxygen-burning
shell visible at roughly $2\,\Msun$ .  In Figure~\ref{fig5}\emph{a}, this shell starts  at approximately
$\log(\mbox{time till core collapse / yr})$\footnote{Here we assume
  that the core collapses in $250$ ms after the end of our pre-SN
  calculation, when the iron core reaches an infall velocity of $1000$
  km/sec} of $-3$ and   mixes with the
carbon-rich layer above, whereas in Figure \ref{fig5}\emph{b} where
$R_{3\alpha}$ is increased by $18\,\%$ this burning layer starts at
$\log(\mbox{time till core collapse / yr})$ of $-3.5$ but does not mix
with the carbon-rich layer above.  In the case of merged layers
$^{60}$Fe is efficiently destroyed whereas it remains unchanged in
the separated carbon layer (Figures~\ref{fig5}\emph{c} and
\ref{fig5}\emph{d}).  $^{44}$Ti, on the other hand, is efficiently
produced at the bottom of the hot merged layer and mixed out
throughout the entire convective domain, whereas in the case of
separated layers only some of it is made at the bottom of the
oxygen-burning layer and little is made in the carbon layer.  The
merging of the layers is also not beneficial to the production of
$^{26}$Al; this works better at the bottom of the separated
carbon-burning layer of the star with the increased $R_{3\alpha}$.

\subsection{$^{26}$Al}

A substantial amount of $^{26}$Al is already present at core helium
ignition as a result of the $^{25}$Mg(p$,\gamma$)$^{26}$Al reaction
occurring during core H burning.  This reaction is essentially the
only means of production of $^{26}$Al, even during the later stages of
stellar evolution.  During core helium burning we see most of
$^{26}$Al in the H envelope with an abundance peak at the location of
the hydrogen-burning shell.  In the helium-burning core $^{26}$Al is
mostly destroyed.  The amount of $^{26}$Al remains essentially
constant until core O depletion and then increases in the later stages
of stellar evolution where $^{26}$Al is made in C shell burning.  The
amount finally ejected after explosion is about an order of magnitude
larger than the amount present at helium ignition. A discussion of the
reactions that create and destroy $^{26}$Al in later burning stages is
given in \cite{lim06}.

\subsection{$^{60}$Fe}

Only a negligible amount of $^{60}$Fe ($\sim 3\E{-15}\,\Msun$) is
present at core helium ignition.  That amount rapidly increases to
$\sim1\E{-7}\,\Msun$ at core helium depletion and gradually builds up
through the later stages until explosion.  Most of it is
found at the lower part of the split He-burning shell and is made
\emph{after} central oxygen depletion.  This last point underscores
the importance of virtually every stage in the star's life for the
production of this isotope.  $^{60}$Fe is mainly produced by neutron
capture on $^{59}$Fe and destroyed by $^{60}$Ni($n,\gamma$)reactions.  When the standard values of the helium burning
rates are used, the pre-SN results show the destruction of some
$^{60}$Fe; when $R_{3\alpha}$ is increased by $18\,\%$, this does not
occur.  This is due to the merging of the oxygen-burning and
carbon-burning layers in the standard case which is not present in the
case with the increased $R_{3\alpha}$ (Figure~\ref{fig5}).  $^{60}$Fe
from the entire carbon-rich convective zone can be mixed to the bottom
of the very hot oxygen layer where it is partially destroyed.  The
$^{60}$Fe present in the He shell is not directly affected by this
destruction, but due to the merging layers, the shell gets less hot
and hence produces less $^{60}$Fe (Figure~\ref{fig5}\emph{e,f}).
There is a significant difference between the amount of $^{60}$Fe
ejected in those two cases: $3.933\E{-5}\,\Msun$ and
$2.201\E{-4}\,\Msun$, respectively, just as a result of variations in
the helium burning rates.

\subsection{$^{44}$Ti}

The amount of $^{44}$Ti present at core helium ignition is minuscule -
$\sim 0.25\,$g.  It increases through the later burning stages, but
remains below $1\E{-8}\,\Msun$ outside of the mass cut until core Si
depletion for the case $R_{3\alpha}=1.18$, and still later for the
standard values of the helium burning rates.  $^{44}$Ti is only made
by O burning and later phases that all start off below the SN mass
cut.  A discussion of the reactions that create and destroy $^{44}$Ti is given in \cite{voc08}. Remarkably, there is a significant amount of $^{44}$Ti at the
pre-SN stage just outside of the Fe-core in the Si-Shell.  All of this
disappears inside the remnant in these two cases; only some $^{44}$Ti
that is made in O shell burning is being ejected.  This contribution
is significantly bigger in the case of the merged O/Ne/C layers of the
standard case as compared to the split layers of the case with
$R_{3\alpha}=1.18$.  Whereas the standard case has 10x more $^{44}$Ti
in the pre-SN model outside the SN mass cut than the
$R_{3\alpha}=1.18$ case, the final production in both cases is
entirely dominated by SN explosions.  It would, therefore, be sensitive to uncertainties in the explosion energy and the asymmetries in the SN, and
how these depend on a pre-SN structure, which is quite different
in the two cases.  That is, if the explosion was  different in the
two cases and not a ``standard'' $1.2\E{51}\,\erg$ explosion
as we have assumed here, the $^{44}$Ti yields might  be
quite different.

\section{Discussion}

We have shown that the uncertainties in the helium burning reaction
rates cause large differences in the predicted yields of $^{26}$Al,
$^{44}$Ti, and $^{60}$Fe, with the changes in $^{60}$Fe being
particularly large.  These yields are also sensitive to uncertainties
in the rates of other important reactions and in the approximate
treatment of convection \citep{lim06,woo07}; there may also be sites
other than massive stars that contribute to the abundances of
$^{26}$Al, $^{44}$Ti, and $^{60}$Fe.  The changes we have examined
here, however, give a lower limit on the overall uncertainties which
is sufficiently large so as to limit deductions that depend on their
yields.  Including other uncertainties will only make the situation
worse.

We considered only stars with masses less than $30\,\Msun$ and
consequently did not include the contributions of Wolf-Rayet stars.
In the models of \cite{lim06}, Wolf-Rayet winds contribute less than
half the yield of $^{26}$Al in these heavier stars; \cite{woo07} find
a similar contribution of wind and explosion for a $60\,\Msun$ star.
It seems reasonable to assume that the explosive process in Wolf-Rayet
stars is subject to uncertainties similar to those we found for
lighter stars, and that their overall contribution will be subject to
similar uncertainties.  This remains to be confirmed by further
calculations.

\begin{figure*}
\centering
\includegraphics[angle=90,width=0.475\textwidth]{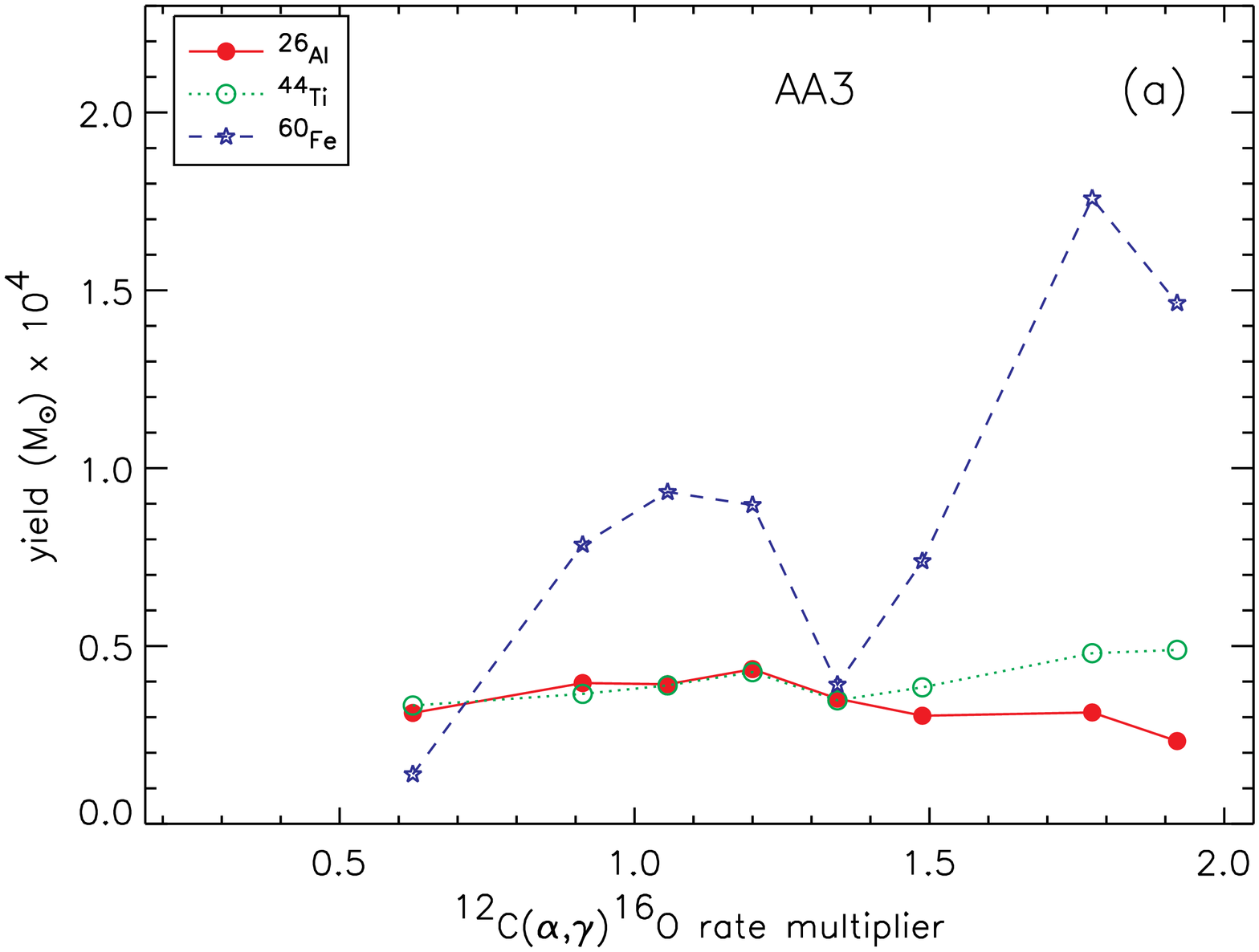}
\hfill
\includegraphics[angle=90,width=0.475\textwidth]{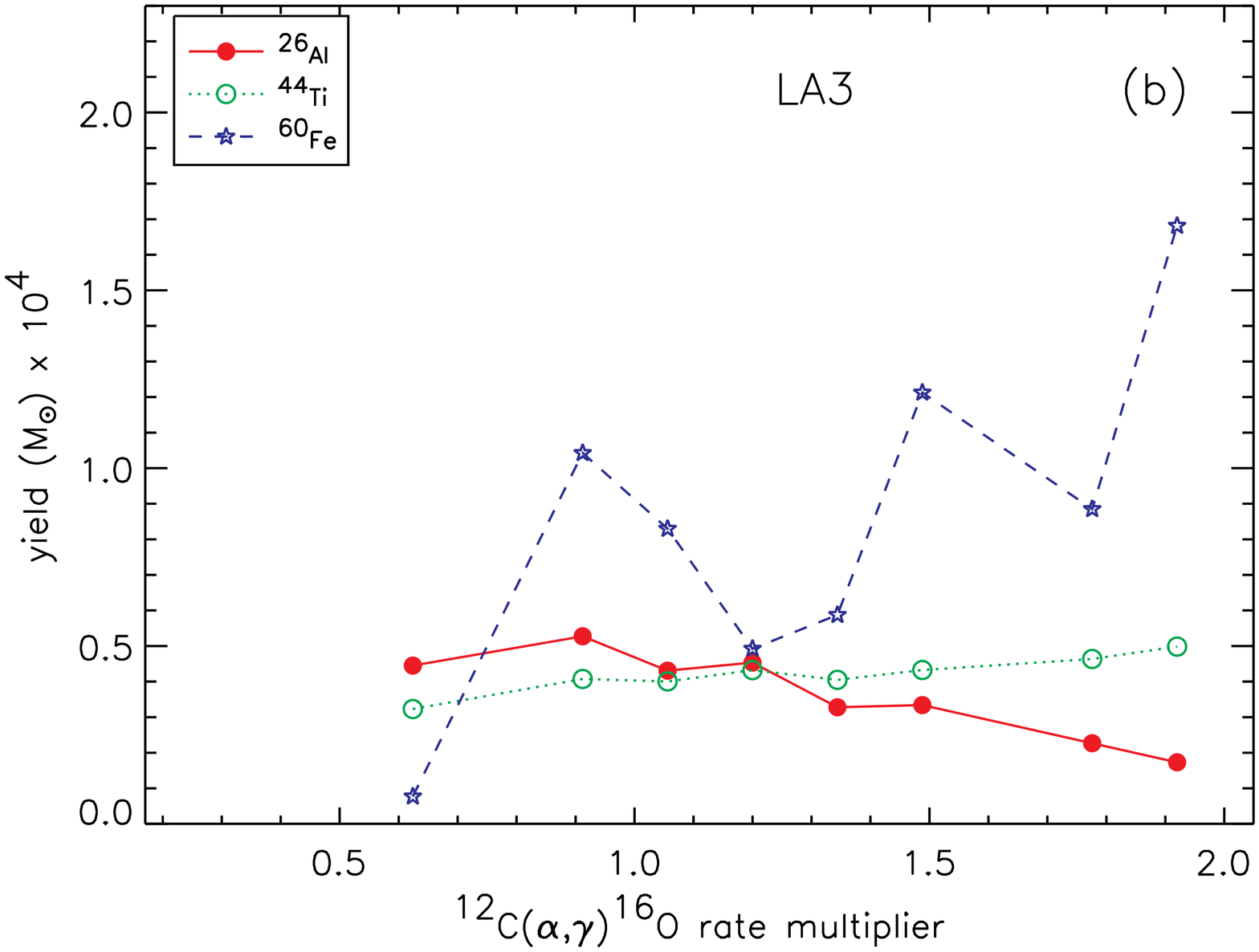}
\hfill
\includegraphics[angle=90,width=0.475\textwidth]{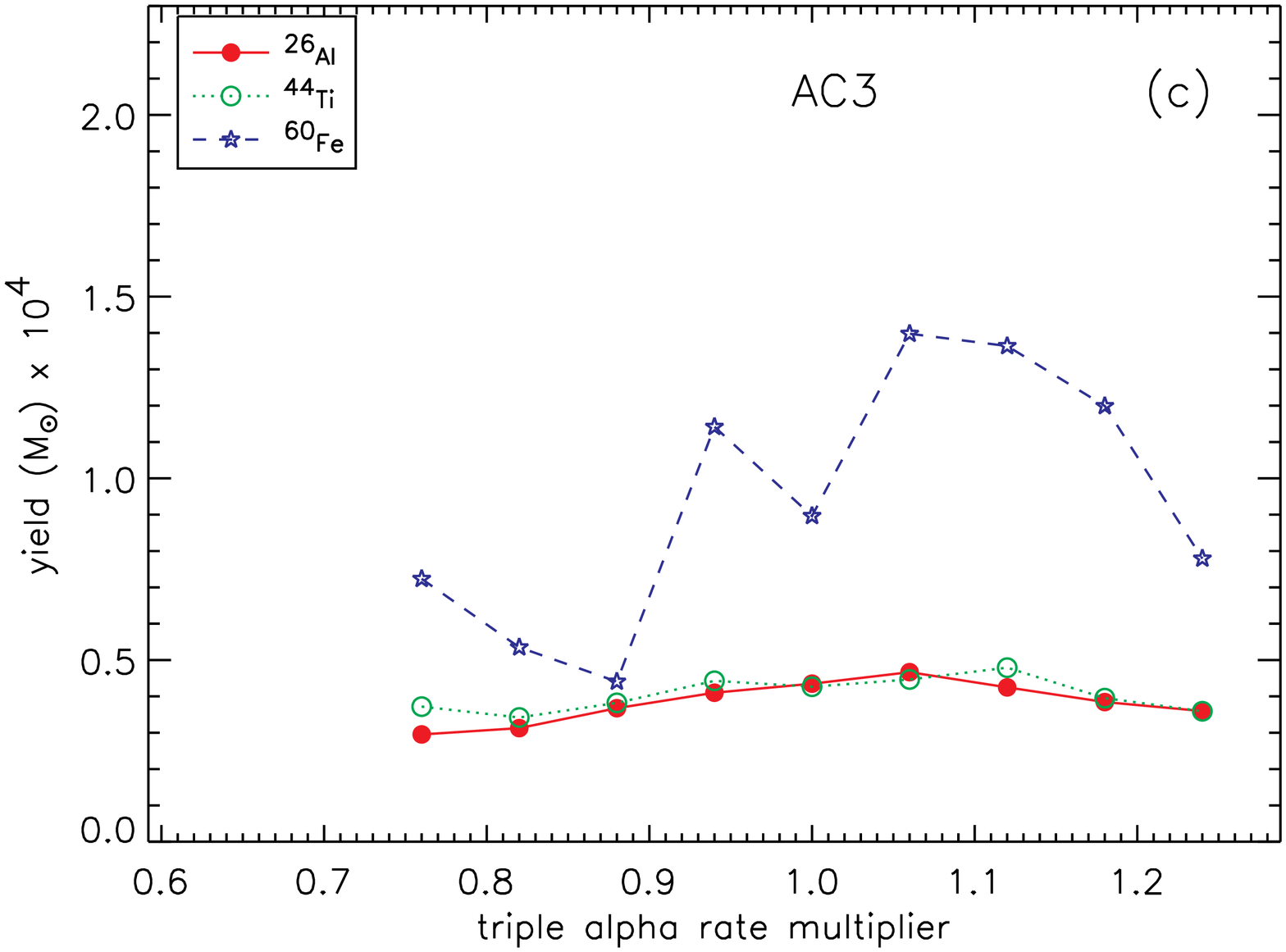}
\hfill
\includegraphics[angle=90,width=0.475\textwidth]{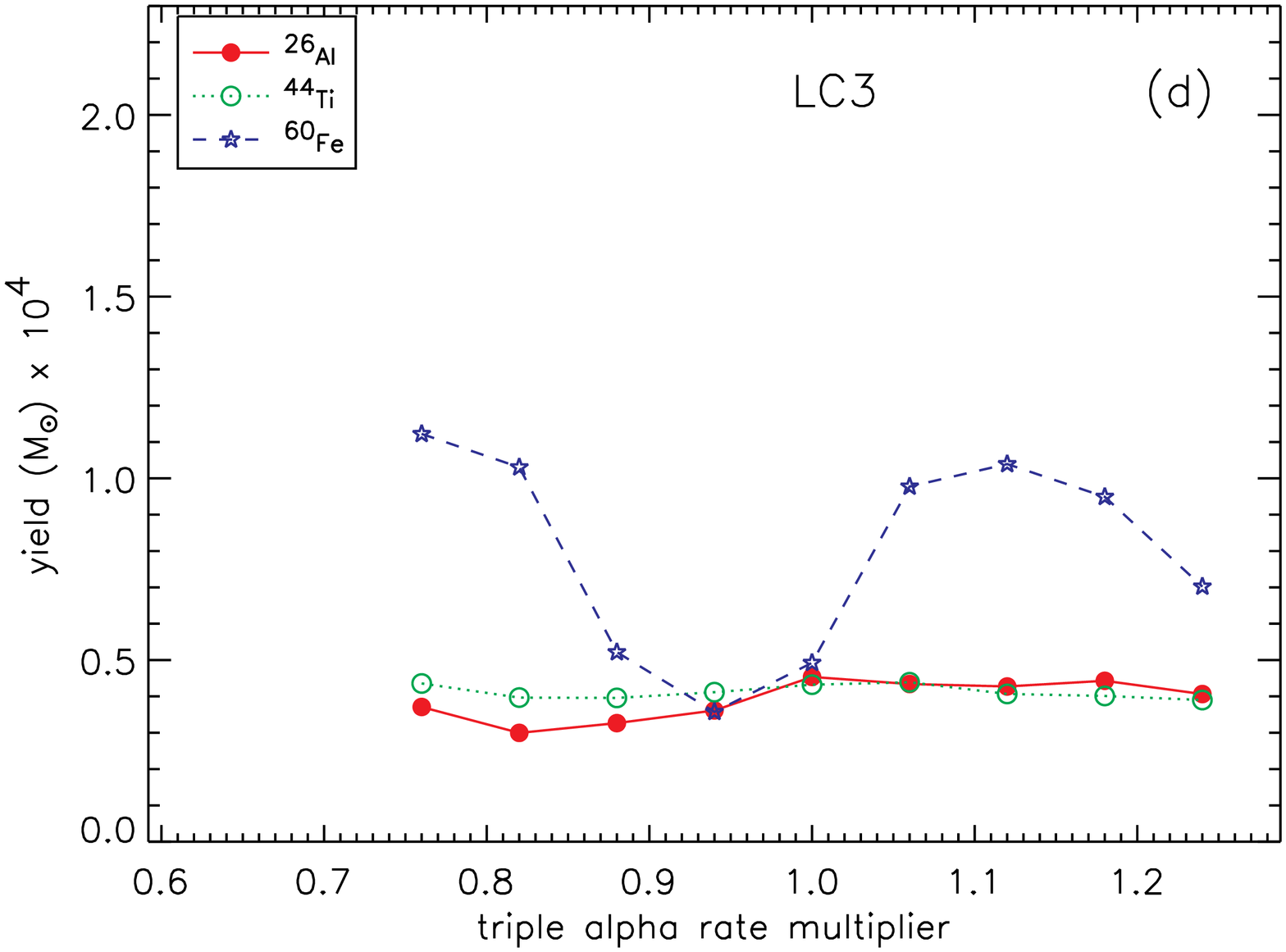}
\hfill
\includegraphics[angle=90,width=0.475\textwidth]{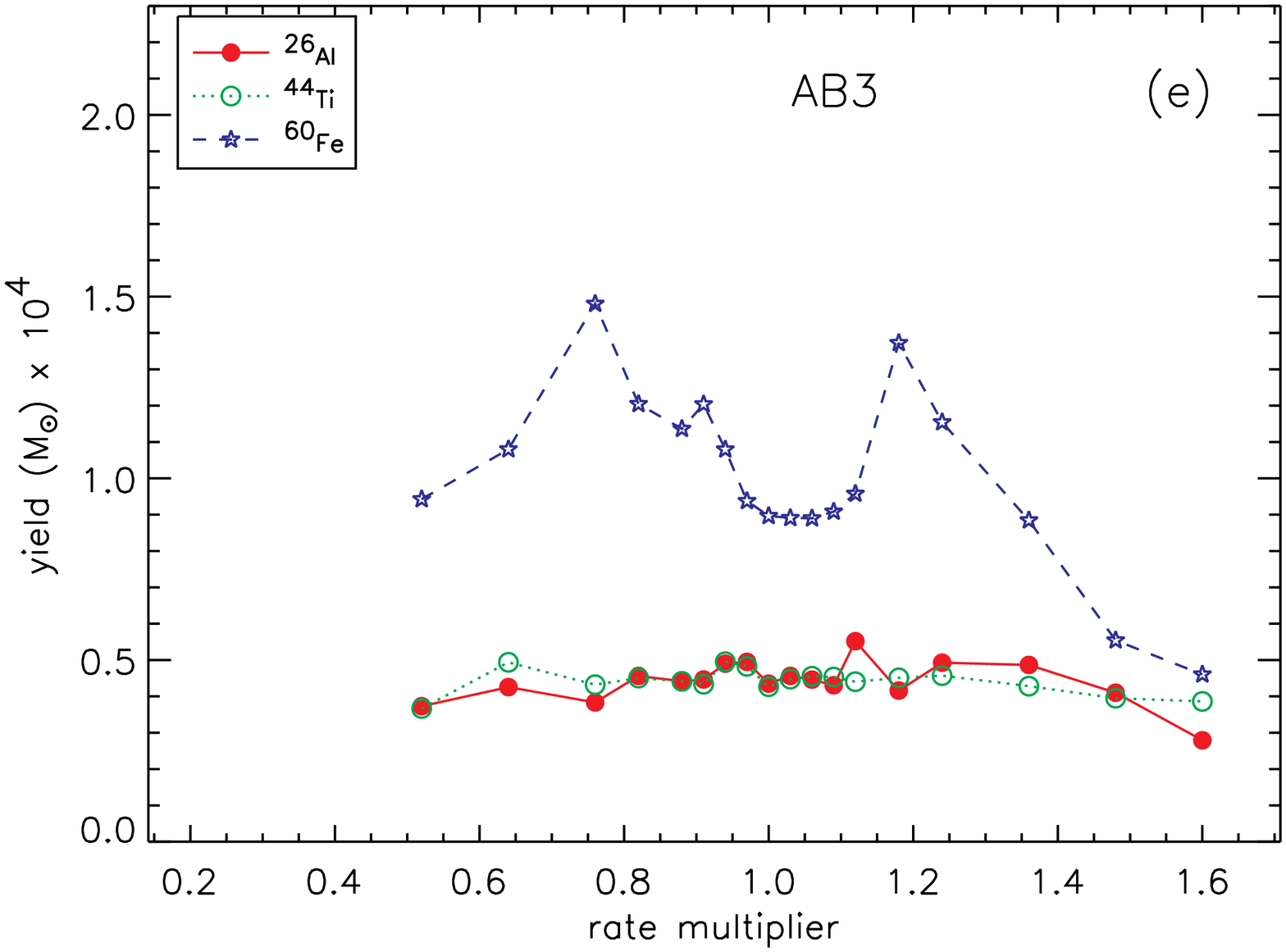}
\hfill
\includegraphics[angle=90,width=0.475\textwidth]{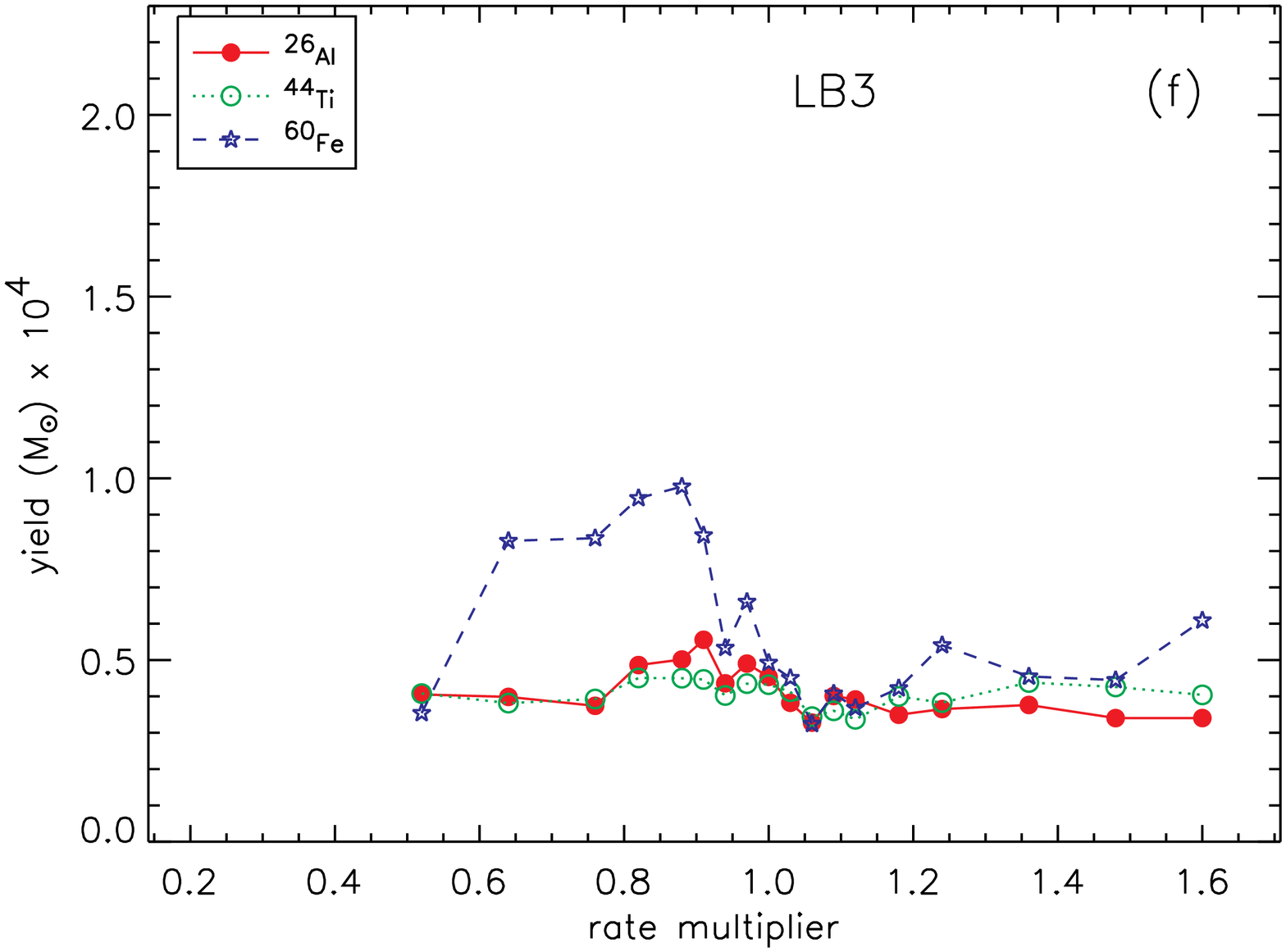}
\caption{Yields versus reaction rate, averaged over 3 stars
  ($15\,\Msun$, $20\,\Msun$, and $25\,\Msun$).  For details see the
  text.  \textbf{(a)} AA series. \textbf{(b)} LA series. \textbf{(c)}
  AC series.  \textbf{(d)} LC series. \textbf{(e)} AB
  series. \textbf{(f)} LB series.}
\label{fig3}
\end{figure*}

\begin{figure*}
\centering
\includegraphics[angle=90,width=0.475\textwidth]{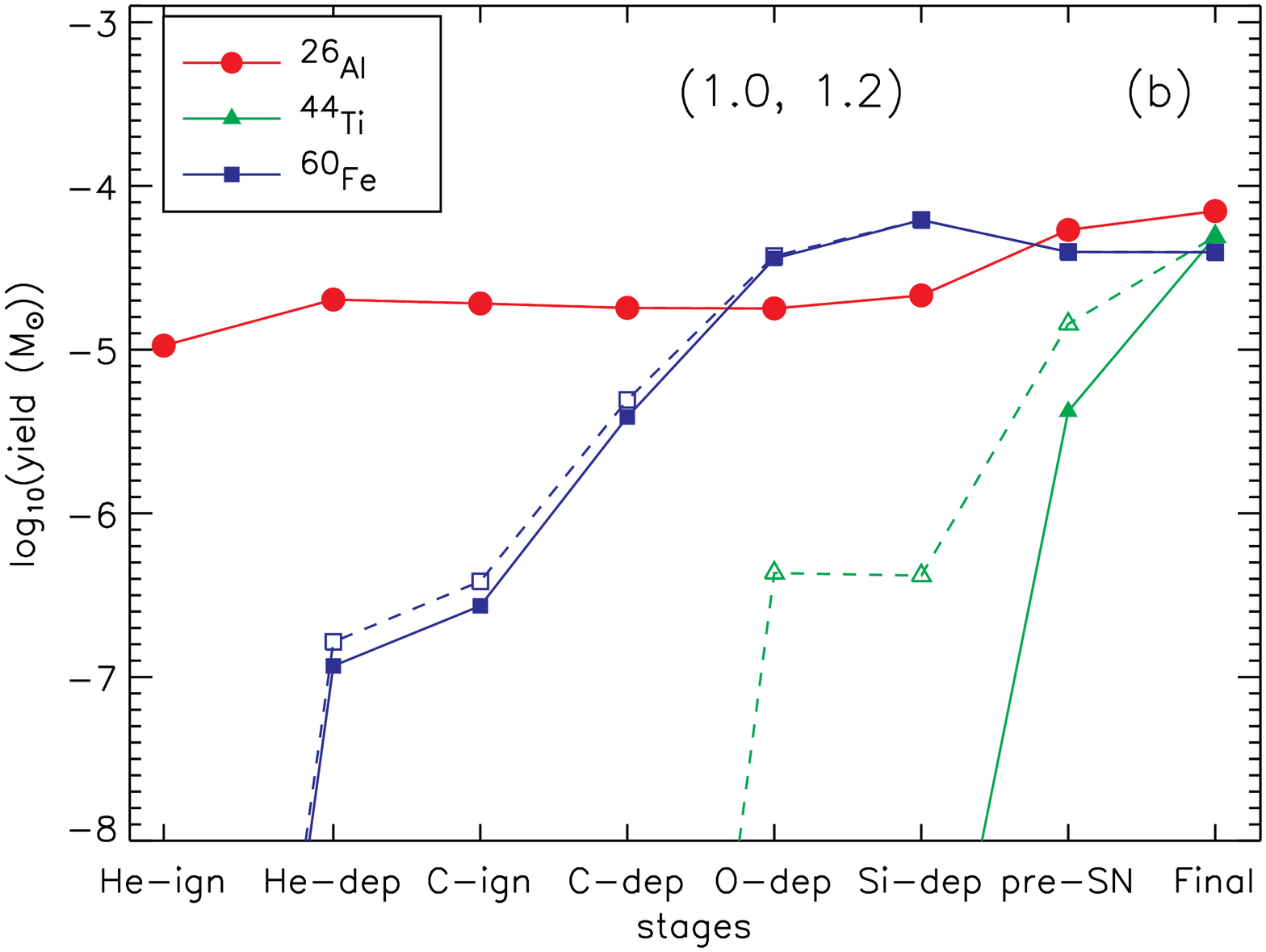}
\hfill
\includegraphics[angle=90,width=0.475\textwidth]{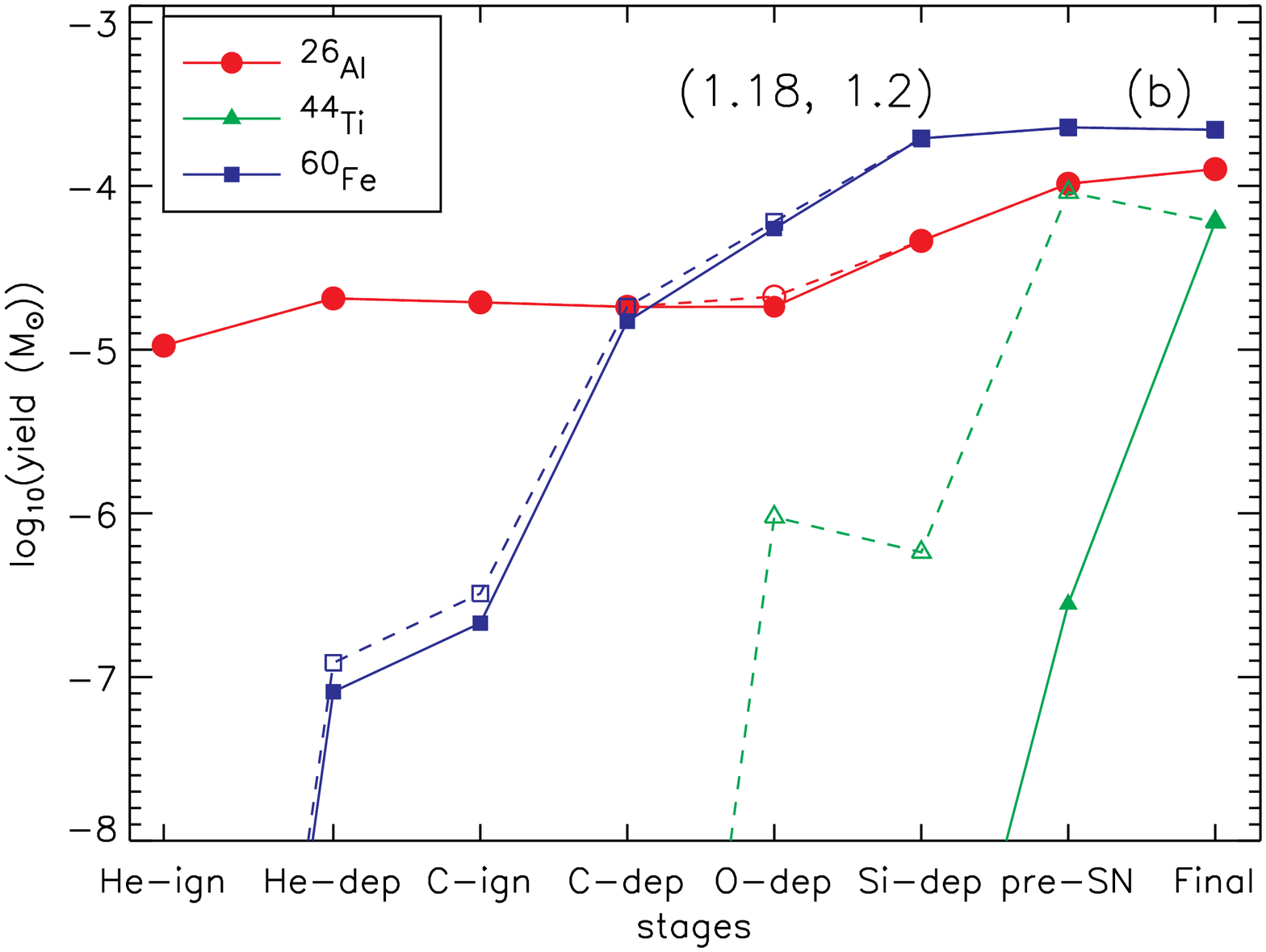}
\caption{%
  Solid symbols and lines give yields (total mass of the isotope
  outside the final mass cut of the SN, see Figure~\ref{fig5}) at
  various evolutionary stages for a $25\,\Msun$ star using the L03
  initial abundances.  The definition of the stages labeled on the
  abscissas are given in the text.  The hollow symbols and lines show
  the total yield inside the star (including wind) as shown in the
  bottom row of Figure~\ref{fig5}.  \textbf{(a)} Standard values of
  the helium burning reaction rates (i.e., multipliers for
  $R_{3\alpha}$ and $R_{\alpha,12}$: $1.0$ and $1.2$,
  respectively). \textbf{(b)} $R_{3\alpha}$ higher by $18\,\%$ (i.e.,
  multipliers for $R_{3\alpha}$ and $R_{\alpha,12}$: $1.18$ and $1.2$,
  respectively).}
\label{fig4}
\end{figure*}

Finally we consider two comparisons of the nucleosynthesis yields and
astrophysical observations.  As was mentioned above, in the steady
state expected for isotope lifetimes short compared to galactic
evolution times, the production rate of gamma rays is proportional to
the amount of material made in a supernova explosion.  \cite{die06b}
have shown that the observed $^{26}$Al gamma ray flux corresponds to a
supernova rate of $1.9\pm1.1$ events/century for an assumed value of
the supernova yield.  Since the gamma ray flux is proportional to the
supernova yield for $^{26}$Al, it is subject to the uncertainties in
the yield shown in Figure \ref{fig3}\emph{a}-\emph{f}; we must
conclude that the uncertainty in this estimate of the supernova rate
is much too small.

A second comparison is to the ratio of the gamma fluxes from $^{26}$Al
and $^{60}$Fe.  If one assumes that for both nuclei the gamma rays
come from supernovae, and that the contributing supernovae have the
same distribution in space, then the ratio of the fluxes is the ratio
of the numbers of nuclei produced, or
$\textrm{flux}(60)/\textrm{flux}(26)=(26/60)\B{\textrm{yield}(60)/\textrm{yield}(26)}$.
The yield ratios are shown in Figure~\ref{fig6}\emph{a}-\emph{d}.
These ratios should be compared to $60/26$ times the flux ratios
summarized in Table~\ref{tbl-2} of \citet{wan07}, namely
$(60/26)(0.15\pm0.06)$.  The ratios shown in Fig.~\ref{fig6} are much
larger for most values of the helium burning reaction rates, although
the smallest values of the $^{12}$C($\alpha,\gamma$)$^{16}$O reaction
rates do give ratios approaching the observed values.  But given the
size of the variations shown, it will not be possible to show that
this value agrees with supernovae predictions until the helium burning
reaction rates are much better known.

A different sort of comparison arises for $^{44}$Ti.  It is seen in
the supernova remnant Cas~A, but not elsewhere.  Given conventional
supernova rates, additional occurrences were expected \citep{the06}.
$^{44}$Ti production is not particularly sensitive to the helium
burning rates--see Figure~\ref{fig3}\emph{a}-\emph{f}.  Possible
explanations include a lower production rate as occurs for some
reaction rate combinations, or a lower supernova rate than assumed.
As was pointed out above, $^{44}$Ti production may  be
particularly sensitive to the supernova explosion physics.
\citet{you07} have shown that these effects can be very large.
A reference to our explosion model including supernova burning
temperatures and time-scales as well as sensitivity of yields to
explosion energy can be found in \citet{woo95,rau02}.

\section{Conclusions}

We explored changes in the core-collapse supernovae yields of
$^{26}$Al, $^{44}$Ti, and $^{60}$Fe arising from variations in the
triple alpha and $^{12}$C($\alpha,\gamma$)$^{16}$O reaction rates.  We
find that these changes are significant, sometimes more than an order
of magnitude from the minimum to the maximum yield within the $\pm
2\,\sigma$ range of rate uncertainty considered here for both
reactions.  The changes were most important for $^{60}$Fe, smaller for
$^{26}$Al, and still smaller for $^{44}$Ti.  The yield-rate
relationship was found to be complex, i.e., non-monotonic, and
sometimes with large changes for small changes in reaction rates.  For
this reason it is difficult to give a simple statistical uncertainty
for the effects of changes in a reaction rate. The high and low
predictions provide a reasonable estimate for a given uncertainty
range.  For the triple alpha reaction $1\,\sigma$ rate changes are $1.0
\pm 0.12$ and for the $^{12}$C($\alpha,\gamma$)$^{16}$O reaction, $1.2
\pm 0.3$.  Thus for a $25\,\Msun$ star, the $^{60}$Fe yield varies
by a factor of $5.0$ ($3.4$) for $\pm 1\,\sigma$ variations of the
triple alpha ($^{12}$C($\alpha,\gamma$)$^{16}$O) rate.  Other ranges
can be determined from Tables~\ref{tbl-2} and \ref{tbl-3}. Obviously the 
detailed dependencies of the isotope production on the rates we varied 
may differ in other model calculations as they also depend on the stellar 
physics and modeling details.

We have isolated the effects on $^{26}$Al, $^{44}$Ti, and $^{60}$Fe
yields, of variations in the helium burning reaction rates within
their experimental uncertainties.  These effects turn out to be large
and limit the conclusions we can draw from comparisons with
observational data.  They provide a lower limit to the total
uncertainties that will remain even if all the uncertainties mentioned
earlier are eliminated.  This highlights the pressing need for
improvements in our knowledge of the triple alpha and
$^{12}$C($\alpha,\gamma$)$^{16}$O reaction rates.

We also explored the dependence of the $^{26}$Al, $^{44}$Ti, and
$^{60}$Fe yields on the initial solar abundances used for the initial
composition of the star: \cite{and89} versus \cite{lod03}.  There
again the dependence is complex, but smaller than that induced by
changes in the helium burning reaction rates.

\acknowledgements

We thank Robert Hoffman for providing the solar abundance files used
in this study and Stan Woosley for helpful discussions, including
studies on the relative influence of the two reaction rates. This
research was supported in part by the US National Science Foundation
grants PHY06-06007 and PHY02-16783, the latter funding the Joint
Institute for Nuclear Astrophysics (JINA), a National Science
Foundation Physics Frontier Center.  AH performed his
contribution under the auspices of the National Nuclear Security
Administration of the US Department of Energy at Los Alamos National
Laboratory under contract DE-AC52-06NA25396, and has been supported by
the DOE Program for Scientific Discovery through Advanced Computing
(SciDAC; DE-FC02-01ER41176), and by the US Department of Energy under
grant DE-FG02-87ER40328.

\clearpage

\begin{deluxetable}{lrrrr}
\tablecolumns{5}
\tablecaption{Yields ($M_{\sun}$)$\times10^5$: standard values of $R_{3\alpha}$ and $R_{\alpha,12}$.\label{tbl-1}}
\tablehead{\colhead{} & \colhead{15 $M_{\sun}$} & \colhead{20 $M_{\sun}$} & \colhead{25 $M_{\sun}$} & \colhead{3 star ave.\tablenotemark{a}}}
\startdata
\\
\multicolumn{5}{c}{AG89 abundances:}\\
\\
$^{26}$Al &	2.789 & 2.934 & 9.781 & 4.349\\
$^{44}$Ti  &	4.221 & 3.333 & 5.707 & 4.266\\
$^{60}$Fe &     6.333 & 4.748 & 20.76 & 8.965\\

\\
\multicolumn{5}{c}{L03 abundances:}\\
\\
$^{26}$Al & 2.443 & 5.962 & 7.023 & 4.535\\	
$^{44}$Ti & 4.263 & 4.039 & 4.846 & 4.322\\	
$^{60}$Fe & 7.481 & 1.727 & 3.933 & 4.924\\
\enddata
\tablenotetext{a}{IMF-averaged over the 15, 20, and 25 $M_{\sun}$ models.}
\end{deluxetable}

\clearpage

\begin{deluxetable}{lrrrrrrrrrrrrrrrr}
\tabletypesize{\scriptsize}
\rotate
\tablecolumns{17} \tablecaption{Yields ($M_{\sun}$) for $^{26}$Al, $^{44}$Ti and $^{60}$Fe as a function of the R$_{3\alpha}$ and R$_{\alpha,12}$ rates for stars of 15 $M_{\sun}$, 20 $M_{\sun}$, and 25 $M_{\sun}$ with AG89
initial abundances\label{tbl-2}}
\tablewidth{0pt}
\tablehead{
\colhead{} &  \multicolumn{3}{c}{15 $M_{\sun}$}   &  \colhead{}   &  \multicolumn{3}{c}{20 $M_{\sun}$}  &  \colhead{}   & \multicolumn{3}{c}{25 $M_{\sun}$}  &  \colhead{}   & \multicolumn{4}{c}{Three Star Average}  \\
\cline{2-4} \cline{6-8} \cline{10-12} \cline{14-17}\\
\colhead{Rate Multiplier\tablenotemark{a}} & \colhead{Al} & \colhead{Ti} & \colhead{Fe} & \colhead{}   & \colhead{Al} & \colhead{Ti} & \colhead{Fe} & \colhead{}   & \colhead{Al} & \colhead{Ti} & \colhead{Fe} &  \colhead{}     & \colhead{Al} & \colhead{Ti} & \colhead{Fe}  & \colhead{Fe/Al} \\
}
\startdata
\\
\multicolumn{17}{c}{R$_{3\alpha}$  varied, R$_{\alpha,12}$  constant}\\
\\
(0.76,1.20)	& 2.31E-05	& 3.25E-05	& 4.32E-05	&	&  3.08E-05	& 3.92E-05	 & 1.20E-04	&     & 4.17E-05	& 4.45E-05	& 6.79E-05	& 	& 2.95E-05	& 3.72E-05	& 7.23E-05	& 2.45E+00	\\	
(0.82,1.20)	& 1.91E-05	& 3.68E-05	& 9.08E-05	& 	&  3.89E-05	& 2.75E-05	 & 8.98E-06	&     & 4.73E-05	& 3.79E-05	& 3.43E-05	& 	& 3.13E-05	& 3.42E-05	& 5.35E-05	& 1.71E+00	\\
(0.88,1.20)	& 2.45E-05	& 3.60E-05	& 5.32E-05	& 	& 2.17E-05	& 3.27E-05	 & 2.33E-05	& 	& 8.45E-05	& 5.09E-05	& 5.39E-05	& 	& 3.68E-05	& 3.82E-05	& 4.41E-05	& 1.20E+00	\\	
(0.94,1.20)	& 2.28E-05	& 4.03E-05	& 6.88E-05	& 	& 4.30E-05	& 4.14E-05	 & 1.64E-04	& 	& 7.75E-05	& 5.70E-05	& 1.42E-04	& 	& 4.10E-05	& 4.43E-05	& 1.14E-04	& 2.79E+00	\\	
(1.00,1.20)	& 2.79E-05	& 4.22E-05	& 6.33E-05	& 	& 2.93E-05	& 3.33E-05	 & 4.75E-05	& 	& 9.78E-05	& 5.71E-05	& 2.08E-04	& 	& 4.35E-05	& 4.27E-05	& 8.97E-05	& 2.06E+00	\\	
(1.06,1.20)	& 2.91E-05	& 4.26E-05	& 9.41E-05	& 	& 2.54E-05	& 3.81E-05	 & 1.27E-04	& 	& 1.16E-04	& 5.86E-05	& 2.59E-04	& 	& 4.67E-05	& 4.47E-05	& 1.40E-04	& 3.00E+00	\\	
(1.12,1.20)	& 2.35E-05	& 3.33E-05	& 1.23E-04	& 	& 4.63E-05	& 6.24E-05	 & 6.52E-05	& 	& 7.89E-05	& 5.96E-05	& 2.68E-04	& 	& 4.25E-05	& 4.79E-05	& 1.36E-04	& 3.21E+00	\\	
(1.18,1.20)	& 2.38E-05	& 3.30E-05	& 1.15E-04	& 	& 4.02E-05	& 4.23E-05	 & 5.36E-05	& 	& 6.80E-05	& 4.96E-05	& 2.25E-04	& 	& 3.84E-05	& 3.95E-05	& 1.20E-04	& 3.12E+00	\\	
(1.24,1.20)	& 2.24E-05	& 2.83E-05	& 1.94E-05	& 	& 4.35E-05	& 4.09E-05	 & 9.33E-05	& 	& 5.53E-05	& 4.56E-05	& 1.85E-04	& 	& 3.60E-05	& 3.59E-05	& 7.79E-05	& 2.17E+00	\\	
													
\\
\multicolumn{17}{c}{R$_{3\alpha}$  constant, R$_{\alpha,12}$  varied}\\
\\
(1.000,0.624)	& 2.68E-05	& 3.77E-05	& 7.40E-06	& 	& 3.30E-05	& 2.81E-05	& 1.33E-05	& 	& 3.80E-05	& 3.11E-05	& 2.88E-05	& 	& 3.12E-05	& 3.33E-05	& 1.39E-05	& 4.47E-01	\\	
(1.000,0.912)	& 2.81E-05	& 3.20E-05	& 7.39E-05	& 	& 4.60E-05	& 3.90E-05	& 2.55E-05	& 	& 5.56E-05	& 4.30E-05	& 1.63E-04	& 	& 3.96E-05	& 3.65E-05	& 7.85E-05	& 1.98E+00	\\	
(1.000,1.056)	& 2.33E-05	& 3.22E-05	& 6.82E-05	& 	& 4.41E-05	& 4.18E-05	& 4.88E-05	& 	& 6.74E-05	& 4.96E-05	& 2.12E-04	& 	& 3.92E-05	& 3.89E-05	& 9.33E-05	& 2.38E+00	\\	
(1.000,1.200)	& 2.79E-05	& 4.22E-05	& 6.33E-05	& 	& 2.93E-05	& 3.33E-05	& 4.75E-05	& 	& 9.78E-05	& 5.71E-05	& 2.08E-04	& 	& 4.35E-05	& 4.27E-05	& 8.97E-05	& 2.06E+00	\\	
(1.000,1.344)	& 2.49E-05	& 3.64E-05	& 5.51E-05	& 	& 4.08E-05	& 2.64E-05	& 8.09E-06	& 	& 4.93E-05	& 4.30E-05	& 4.90E-05	& 	& 3.52E-05	& 3.47E-05	& 3.92E-05	& 1.11E+00	\\	
(1.000,1.488)	& 2.43E-05	& 3.43E-05	& 5.08E-05	& 	& 3.11E-05	& 3.95E-05	& 1.27E-04	& 	& 4.27E-05	& 4.56E-05	& 4.87E-05	& 	& 3.04E-05	& 3.84E-05	& 7.39E-05	& 2.43E+00	\\	
(1.000,1.776)	& 1.43E-05	& 4.19E-05	& 1.87E-04	& 	& 2.65E-05	& 4.65E-05	& 1.40E-04	& 	& 7.61E-05	& 6.37E-05	& 2.02E-04	& 	& 3.13E-05	& 4.80E-05	& 1.76E-04	& 5.61E+00	\\	
(1.000,1.920)	& 1.42E-05	& 4.24E-05	& 1.72E-04	& 	& 1.48E-05	& 4.60E-05	& 8.87E-05	& 	& 5.51E-05	& 6.70E-05	& 1.72E-04	& 	& 2.33E-05	& 4.89E-05	& 1.46E-04	& 6.29E+00	\\	
											
\enddata \tablenotetext{a}{The rate multipliers are given in the
  format (R$_{3\alpha}$ multiplier,R$_{\alpha,12}$ multiplier), where
  the rates being multiplied are the standard rate values chosen as
  described in the text}
\end{deluxetable}

\clearpage

\begin{deluxetable}{lrrrrrrrrrrrrrrrr}
\tabletypesize{\scriptsize}
\rotate
\tablecolumns{17} \tablecaption{Yields ($M_{\sun}$) for $^{26}$Al, $^{44}$Ti and $^{60}$Fe as a function of the R$_{3\alpha}$ and R$_{\alpha,12}$ rates for stars of 15 $M_{\sun}$, 20 $M_{\sun}$, and 25 $M_{\sun}$ with L03
initial abundances\label{tbl-3}}
\tablewidth{0pt}
\tablehead{
\colhead{} &  \multicolumn{3}{c}{15 $M_{\sun}$}   &  \colhead{}   &  \multicolumn{3}{c}{20 $M_{\sun}$}  &  \colhead{}   & \multicolumn{3}{c}{25 $M_{\sun}$}  &  \colhead{}   & \multicolumn{4}{c}{Three Star Average}  \\
\cline{2-4} \cline{6-8} \cline{10-12} \cline{14-17}\\
\colhead{Rate Multiplier\tablenotemark{a}} & \colhead{Al} & \colhead{Ti} & \colhead{Fe} & \colhead{}   & \colhead{Al} & \colhead{Ti} & \colhead{Fe} & \colhead{}   & \colhead{Al} & \colhead{Ti} & \colhead{Fe} &  \colhead{}     & \colhead{Al} & \colhead{Ti} & \colhead{Fe}  & \colhead{Fe/Al} \\
}
\startdata
\\
\multicolumn{17}{c}{R$_{3\alpha}$  varied, R$_{\alpha,12}$  constant}\\
\\
(0.76,1.20) & 1.44E-05 & 3.28E-05 & 3.97E-05 &  & 3.70E-05 & 5.00E-05 & 2.39E-04 &  & 8.620-05 & 5.74E-05 & 9.15E-05 &  & 3.71E-05 & 4.36E-05 & 1.12E-04 & 3.03E+00	\\	
(0.82,1.20) & 1.30E-05 & 3.29E-05 & 1.00E-04 &  & 2.94E-05 & 4.22E-05 & 1.36E-04 &  & 6.81E-05 & 5.11E-05 & 6.27E-05 &  & 3.00E-05 & 3.97E-05 & 1.03E-04 & 3.44E+00	\\	
(0.88,1.20) & 1.80E-05 & 3.60E-05 & 4.00E-05 &  & 2.83E-05 & 3.80E-05 & 7.28E-05 &  & 7.13E-05 & 4.97E-05 & 4.98E-05 &  & 3.26E-05 & 3.96E-05 & 5.22E-05 & 1.60E+00	\\	
(0.94,1.20) & 1.90E-05 & 3.71E-05 & 4.00E-05 &  & 3.99E-05 & 5.10E-05 & 3.40E-05 &  & 6.88E-05 & 3.65E-05 & 2.88E-05 &  & 3.61E-05 & 4.12E-05 & 3.57E-05 & 9.89E-01	\\	
(1.00,1.20) & 2.44E-05 & 4.26E-05 & 7.48E-05 &  & 5.96E-05 & 4.04E-05 & 1.73E-05 &  & 7.02E-05 & 4.85E-05 & 3.93E-05 &  & 4.54E-05 & 4.32E-05 & 4.92E-05 & 1.09E+00	\\	
(1.06,1.20) & 2.45E-05 & 3.97E-05 & 8.71E-05 &  & 4.21E-05 & 4.04E-05 & 7.63E-05 &  & 8.64E-05 & 5.79E-05 & 1.51E-04 &  & 4.34E-05 & 4.39E-05 & 9.77E-05 & 2.25E+00	\\	
(1.12,1.20) & 2.33E-05 & 3.73E-05 & 9.44E-05 &  & 2.54E-05 & 3.37E-05 & 5.34E-05 &  & 1.10E-04 & 5.81E-05 & 1.98E-04 &  & 4.27E-05 & 4.07E-05 & 1.04E-04 & 2.43E+00	\\	
(1.18,1.20) & 2.14E-05 & 3.22E-05 & 3.11E-05 &  & 2.23E-05 & 3.89E-05 & 1.06E-05 &  & 1.27E-04 & 5.95E-05 & 2.20E-04 &  & 4.43E-05 & 4.01E-05 & 9.49E-05 & 2.14E+00	\\	
(1.24,1.20) & 1.86E-05 & 2.86E-05 & 1.07E-05 &  & 4.15E-05 & 4.00E-05 & 4.43E-05 &  & 8.87E-05 & 6.08E-05 & 2.40E-04 &  & 4.06E-05 & 3.90E-05 & 7.02E-05 & 1.73E+00	\\	
										
\\						
\multicolumn{17}{c}{R$_{3\alpha}$  constant, R$_{\alpha,12}$  varied}\\
\\
(1.000,0.624) & 2.19E-05 & 2.74E-05 & 4.51E-06 &  & 3.15E-05 & 2.70E-05 & 1.22E-05 &  & 1.11E-04 & 4.91E-05 & 8.29E-06 &  & 4.45E-05 & 3.23E-05 & 7.71E-06 & 1.73E-01	\\			
(1.000,0.912) & 2.72E-05 & 3.24E-05 & 3.87E-05 &  & 5.19E-05 & 4.10E-05 & 2.77E-05 &  & 1.10E-04 & 5.84E-05 & 3.59E-04 &  & 5.27E-05 & 4.07E-05 & 1.04E-04 & 1.98E+00	\\			
(1.000,1.056) & 2.14E-05 & 3.22E-05 & 2.43E-05 &  & 2.24E-05 & 3.87E-05 & 9.25E-05 &  & 1.21E-04 & 5.94E-05 & 1.99E-04 &  & 4.31E-05 & 4.01E-05 & 8.30E-05 & 1.93E+00	\\			
(1.000,1.200) & 2.44E-05 & 4.26E-05 & 7.48E-05 &  & 5.96E-05 & 4.04E-05 & 1.73E-05 &  & 7.02E-05 & 4.85E-05 & 3.93E-05 &  & 4.53E-05 & 4.32E-05 & 4.92E-05 & 1.09E+00	\\			
(1.000,1.344) & 1.78E-05 & 3.62E-05 & 4.20E-05 &  & 2.82E-05 & 3.92E-05 & 8.94E-05 &  & 7.25E-05 & 5.19E-05 & 5.24E-05 &  & 3.28E-05 & 4.05E-05 & 5.87E-05 & 1.79E+00	\\			
(1.000,1.488) & 1.32E-05 & 3.50E-05 & 3.86E-05 &  & 3.42E-05 & 4.72E-05 & 2.63E-04 &  & 7.61E-05 & 5.56E-05 & 1.02E-04 &  & 3.34E-05 & 4.33E-05 & 1.21E-04 & 3.63E+00	\\			
(1.000,1.776) & 6.82E-06 & 3.52E-05 & 1.73E-05 &  & 1.29E-05 & 4.76E-05 & 9.11E-05 &  & 7.21E-05 & 6.92E-05 & 2.45E-04 &  & 2.27E-05 & 4.64E-05 & 8.85E-05 & 3.90E+00	\\			
(1.000,1.920) & 1.14E-05 & 3.92E-05 & 1.92E-04 &  & 1.31E-05 & 4.86E-05 & 7.82E-05 &  & 3.65E-05 & 7.48E-05 & 2.44E-04 &  & 1.73E-05 & 4.98E-05 & 1.68E-04 & 9.72E+00	\\

\enddata \tablenotetext{a}{The rate multipliers are given in the
  format (R$_{3\alpha}$ multiplier,R$_{\alpha,12}$ multiplier), where
  the rates being multiplied are the standard rate values chosen as
  described in the text}
\end{deluxetable}

\clearpage

\newcounter{fignum}
\setcounter{fignum}{\thefigure}

\begin{figure*}
\centering
\includegraphics[angle=90,width=0.475\textwidth]{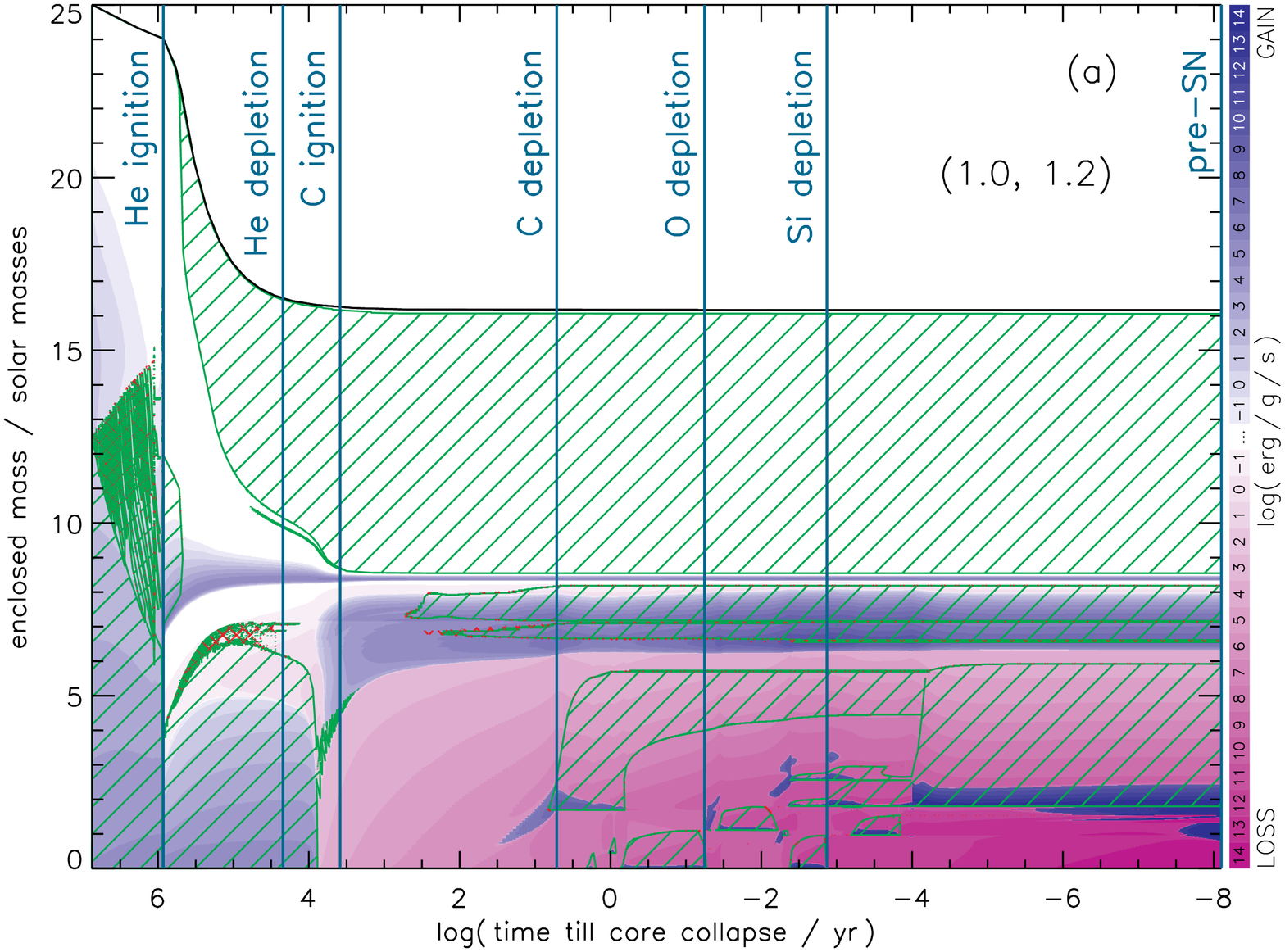}
\hfill
\includegraphics[angle=90,width=0.475\textwidth]{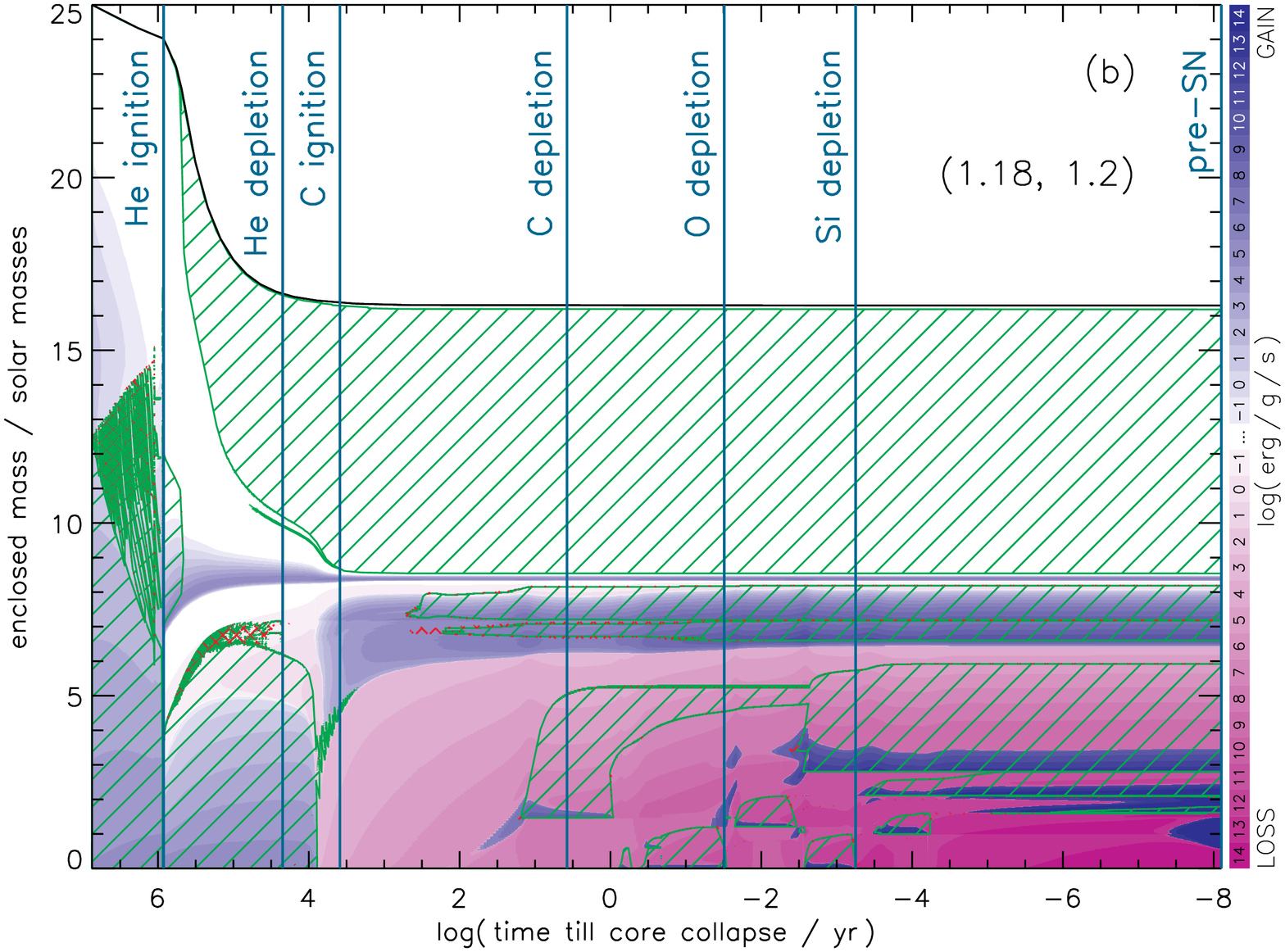}

\bigskip

\includegraphics[angle=90,width=0.475\textwidth]{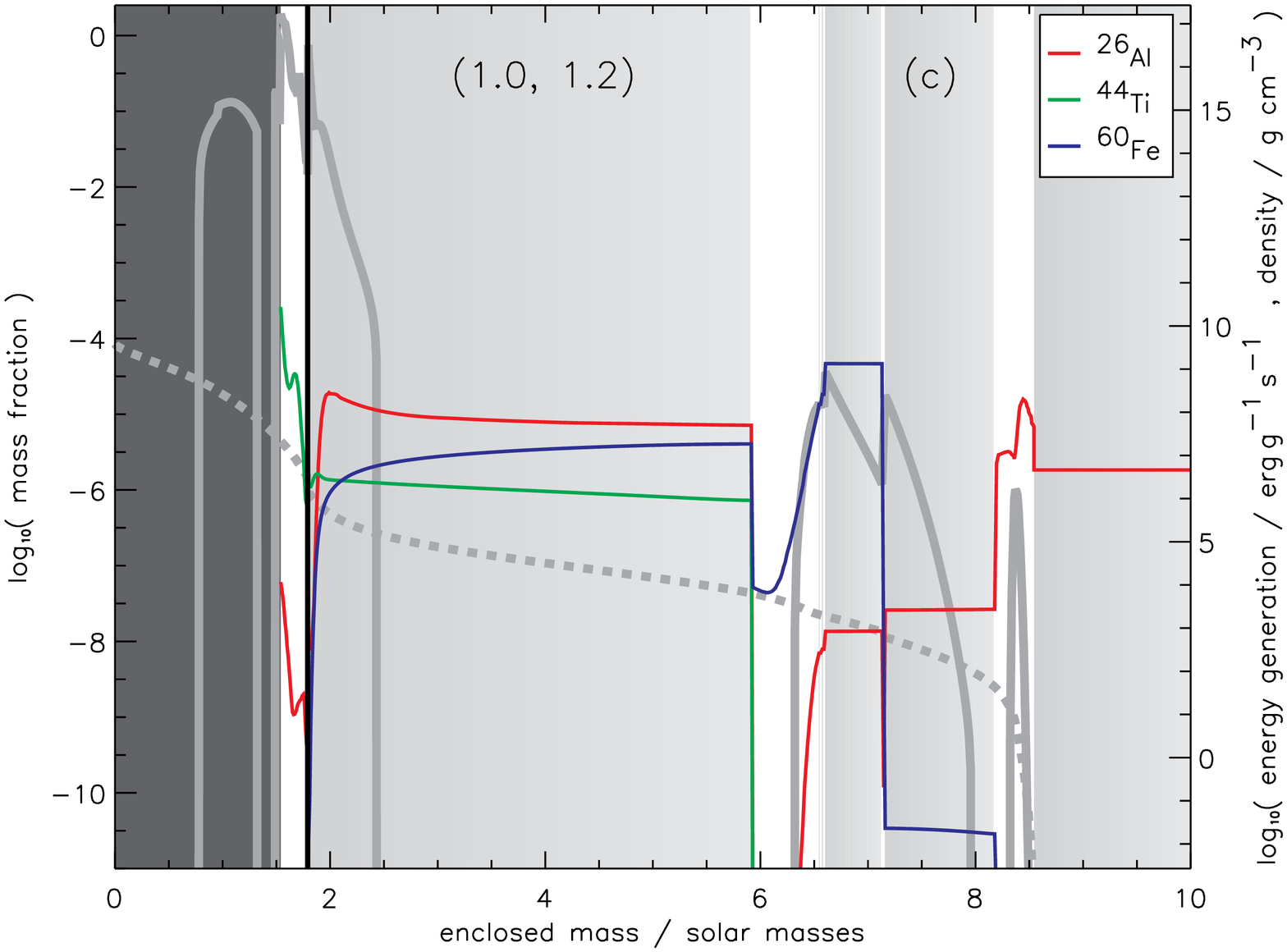}
\hfill
\includegraphics[angle=90,width=0.475\textwidth]{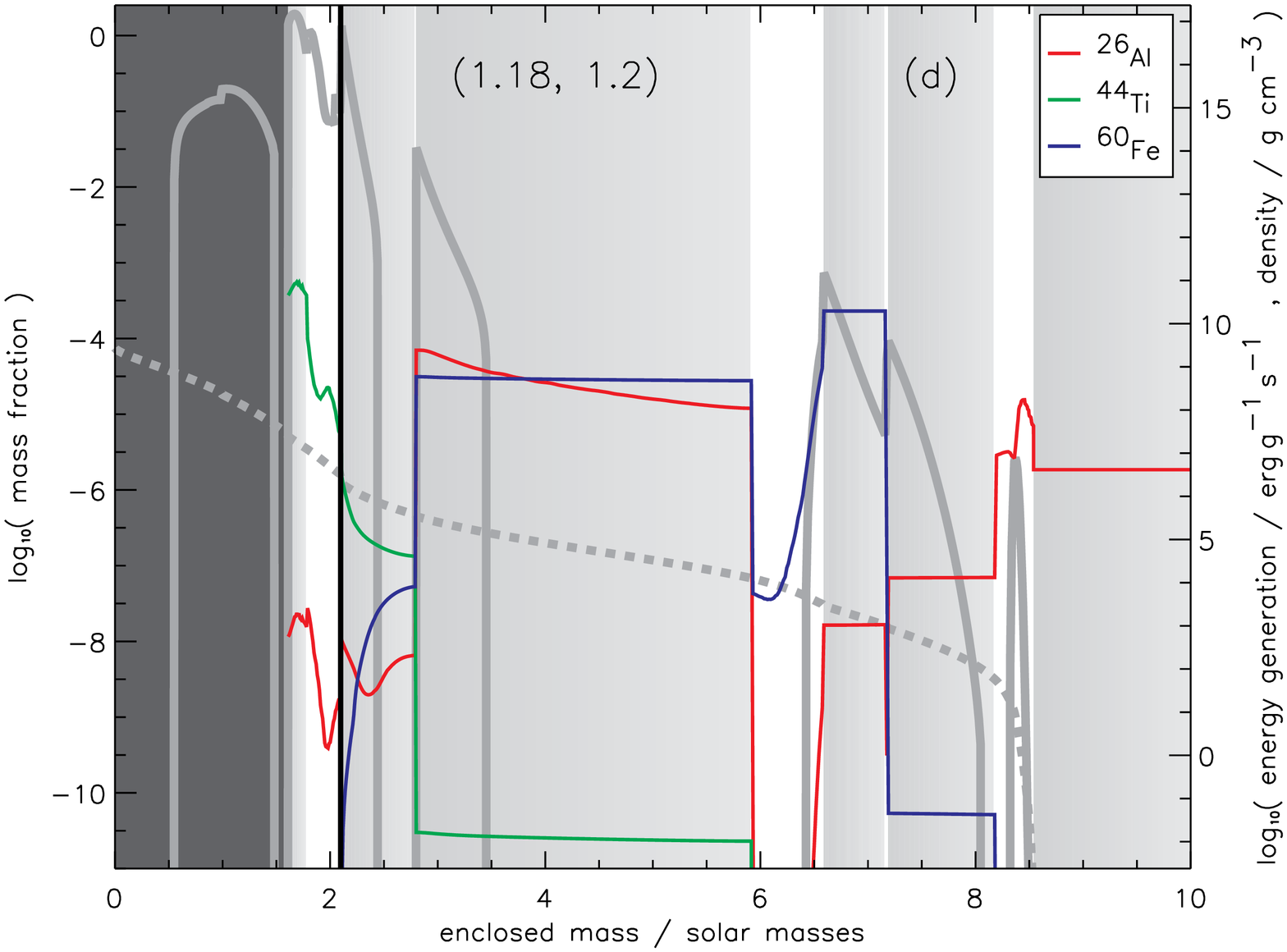}

\bigskip

\includegraphics[angle=90,width=0.475\textwidth]{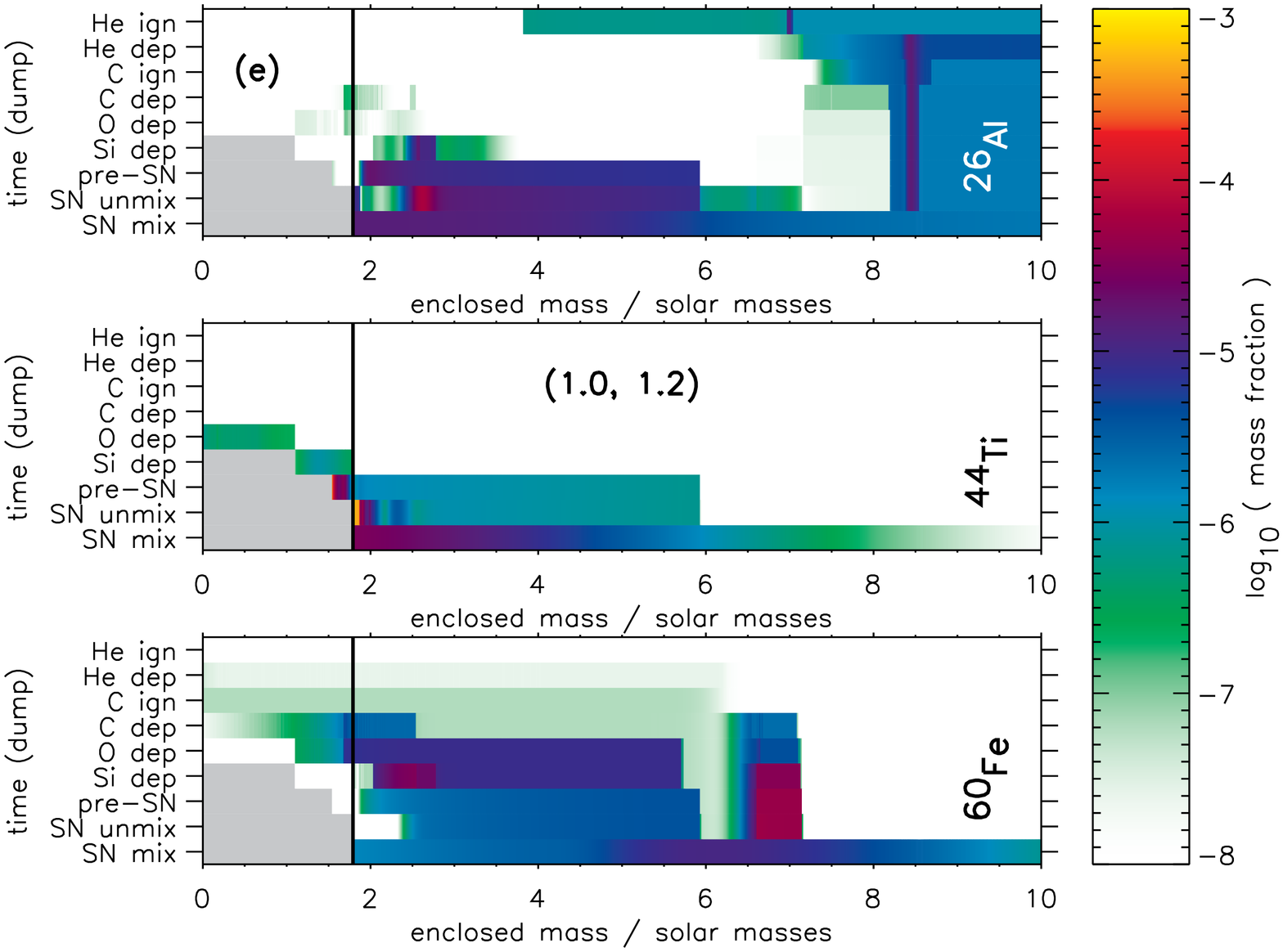}
\hfill
\includegraphics[angle=90,width=0.475\textwidth]{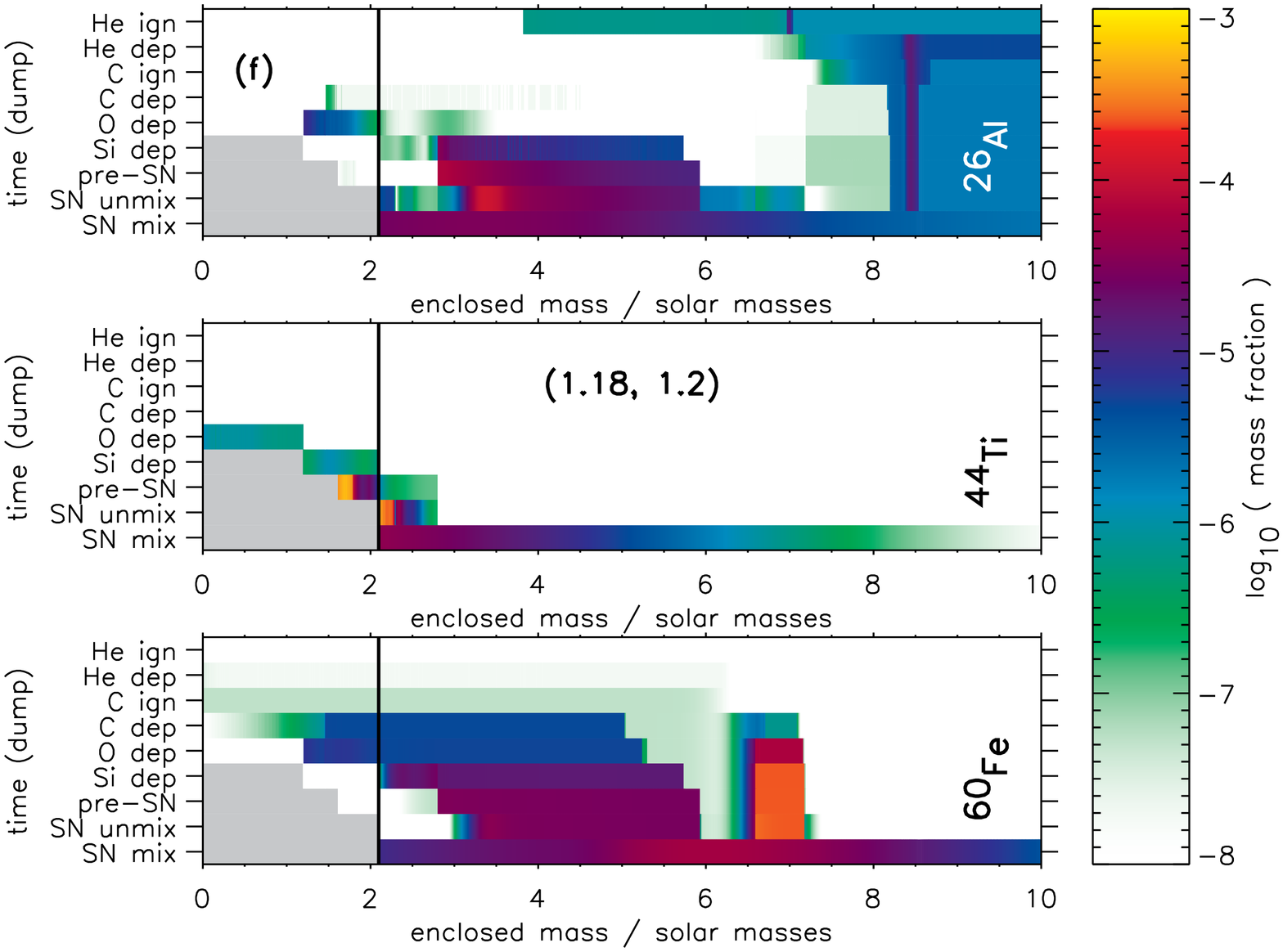}

\end{figure*}

\setcounter{fignum}{\thefigure}

\begin{figure*}
  \caption{%
    Evolution of and isotope distribution in $25\,\Msun$ stars of L03
    abundances.  \textit{Left Column:} Standard values of the helium
    burning reaction rates, $(R_{3\alpha},R_{\alpha,12})=(1.0, 1.2)$.
    \textsl{Right Column:} $R_{3\alpha}$ enhanced by $18\,\%$:
    $(R_{3\alpha},R_{\alpha,12})=(1.18, 1.2)$.  \textbf{Top Row:}
    Convective history as a function of time until core collapse.  The
    ordinate is the mass coordinate from the center of the star.  The
    green hatched areas are fully convective, and the red cross
    hatched areas are semiconvective.  The blue and purple shading
    indicate net energy generation from burning and neutrino losses,
    with blue positive and purple negative.  For more details see
    \cite{woo02}.  The vertical lines show the snapshots we use and
    are explained in more detail in the test.  \textbf{Middle Row:}
    Stellar structure at the pre-supernova stage as a function of the
    enclosed mass (truncated at $10\,\Msun$).  \textsl{Dark gray}
    indicates the neutronized regions in which nucleosynthesis was no
    longer followed.  The \textsl{vertical black line} shows
    the final mass cut of the SN.
    The \textsl{light gray} indicates convective regions.  The
    \textsl{solid gray line} shows the nuclear energy generation and
    the \textsl{dashed gray line} shows the density.  \textbf{Bottom
      Row:} Distribution of isotopes inside the star as a function of
    mass coordinate for different evolution stages ($y$-axis, time
    increased downward).  \textsl{Gray shading} indicates neutronized
    regions (until pre-SN) or location of the SN piston (SN lines) in
    which nucleosynthesis was no longer followed as those regions
    become part of the remnant.  The \textsl{solid black line} shows
    the final mass cut of the SN.  For the two stars shown here there
    was no fallback, so the mass cut and the piston mass coincide;
    this is not the case in general.  }
\label{fig5}
\end{figure*}

\begin{figure*}
\centering
\includegraphics[angle=90,width=0.475\textwidth]{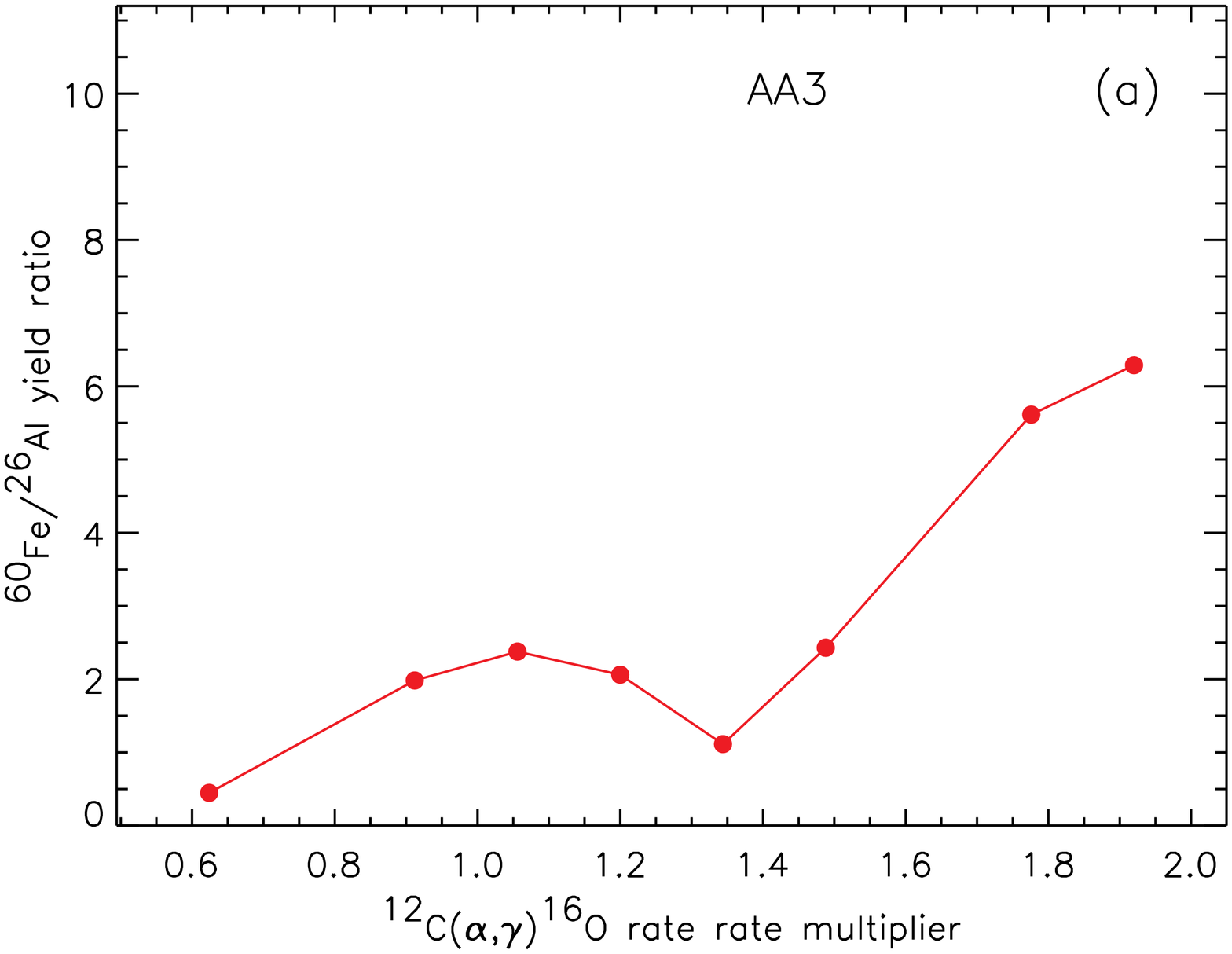}
\hfill
\includegraphics[angle=90,width=0.475\textwidth]{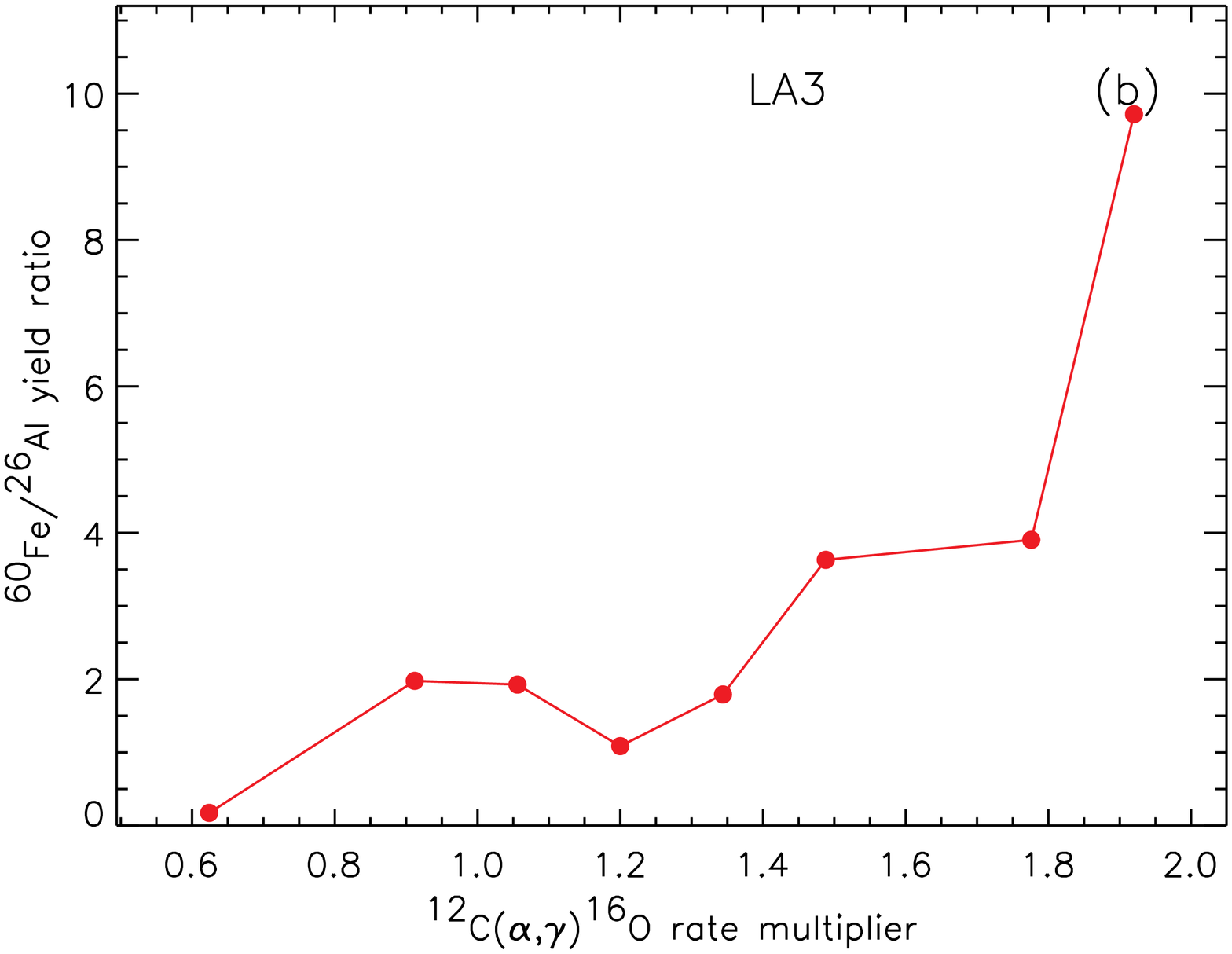}
\hfill
\includegraphics[angle=90,width=0.475\textwidth]{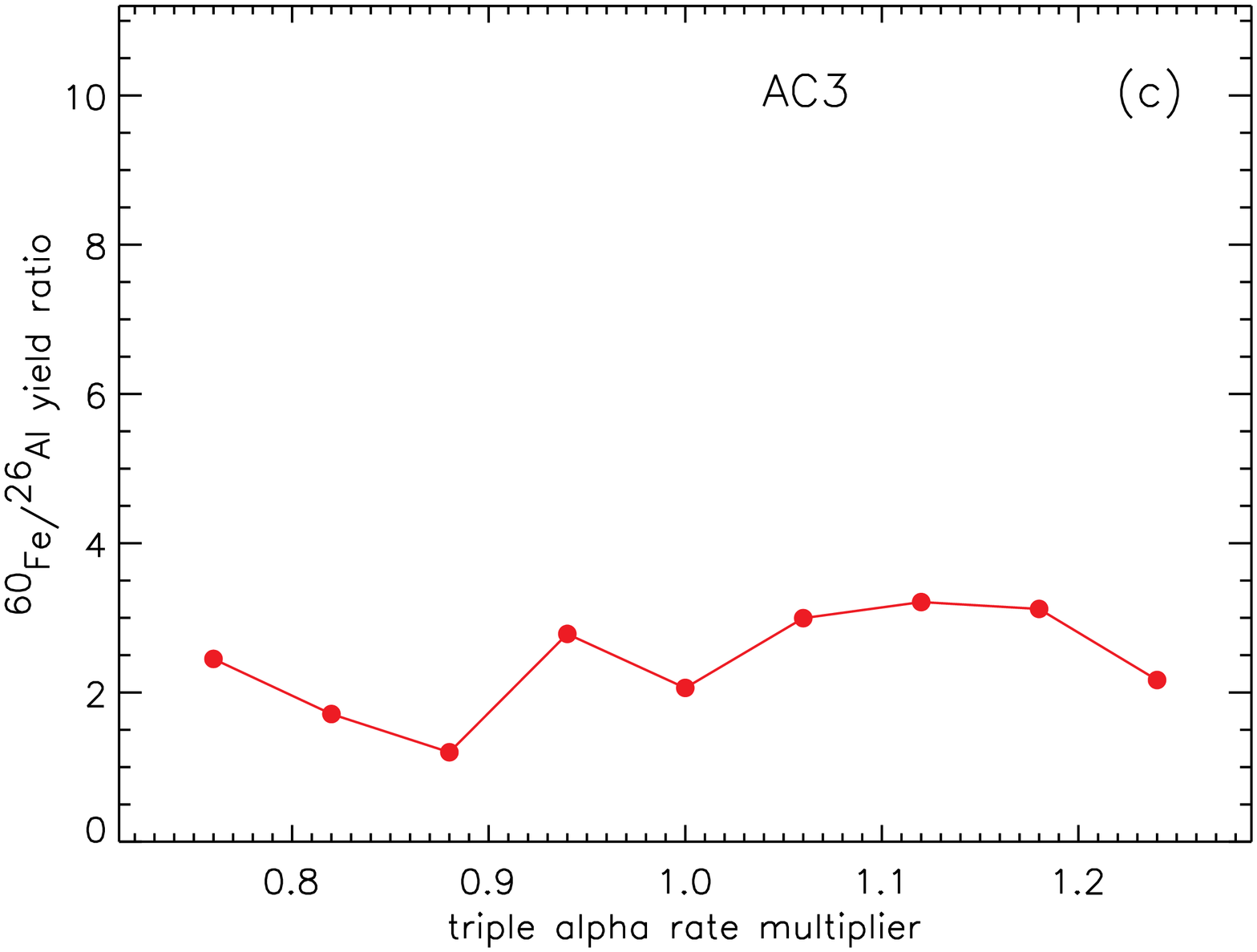}
\hfill
\includegraphics[angle=90,width=0.475\textwidth]{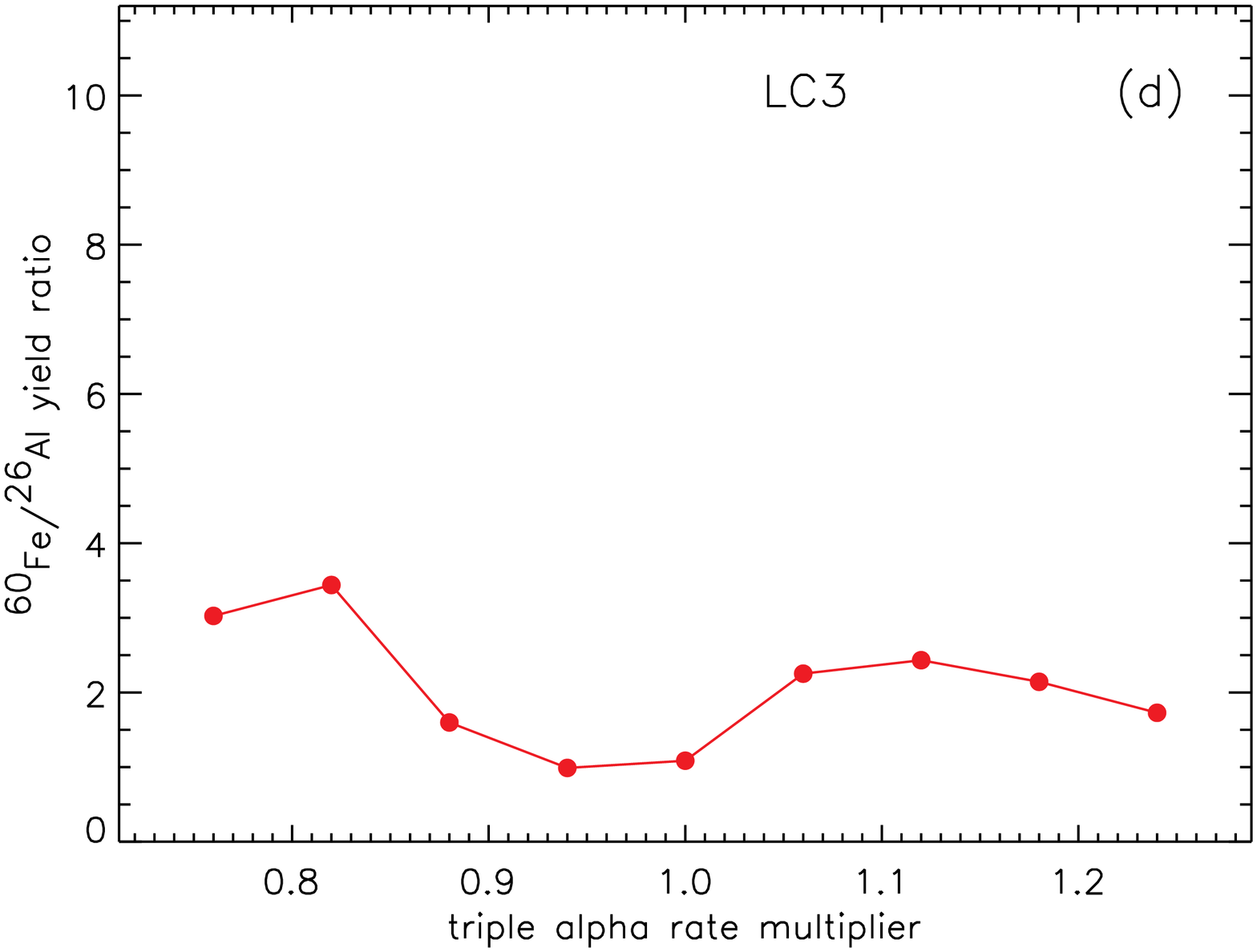}
\caption{Mass ratio of yields of $^{60}$Fe and $^{26}$Al versus
  reaction rate, averaged over three stars ($15\,\Msun$, $20\,\Msun$,
  and $25\,\Msun$).  For details see the text.  \textbf{(a)} AA
  series. \textbf{(b)} LA series. \textbf{(c)} AC series.
  \textbf{(d)} LC series.}
\label{fig6}
\end{figure*}

\end{document}